\documentclass[12pt,letterpaper]{article}
\usepackage[utf8]{inputenc}

\usepackage{graphicx,array}
\usepackage{url}
\usepackage{color}
\usepackage{latexsym}
\usepackage{amsthm}
\usepackage{amsmath}
\usepackage{amssymb}
\usepackage{amsfonts}
\usepackage[numbers,sort&compress]{natbib}
\usepackage{bm}
\usepackage{bbm}
\usepackage{slashed}
\usepackage{mathrsfs}
\usepackage{enumerate}
\usepackage{tikz}
\usepackage{siunitx}
\usepackage{mdframed}
\usepackage{setspace}  
\usepackage{esvect}
\usepackage{physics}
\usepackage{cancel}
\usepackage{subfigure}
\usepackage{enumitem}
\usepackage{booktabs}
\usepackage{tcolorbox}%

\numberwithin{equation}{section}

\usepackage{hyperref} 
\hypersetup{
    colorlinks=true,       
    linkcolor=red,          
    citecolor=blue,        
    filecolor=magenta,      
    urlcolor=blue           
}
\usepackage[all]{hypcap} 
\usepackage{multirow}
\usepackage{multicol}

\newcolumntype{M}[1]{>{\centering\arraybackslash}p{#1}}

\usepackage{natbib}
\setlist[description]{leftmargin=\parindent,labelindent=\parindent}

\usepackage[normalem]{ulem} 

\newcommand{\nc}{\newcommand}

\nc{\beq}{\begin{equation}}
\nc{\eeq}{\end{equation}}
\nc{\beqa}{\begin{eqnarray}}  
\nc{\eeqa}{\end{eqnarray}}  
\nc{\bit}{\begin{itemize}}  
\nc{\eit}{\end{itemize}}  

\newcommand{\eg}{{\it e.g.}}

\usepackage{floatrow}
\newfloatcommand{capbtabbox}{table}[][\FBwidth]

\usepackage{blindtext}

\allowdisplaybreaks

\title{
{\bf Toponia at the HL-LHC and FCC-ee
}
\author{\large Yang Bai$^{\,a,b}$, Ting-Kuo Chen$^{a}$, and Yiming Yang$^{a}$}
\date{\small \it 
$^a$Department of Physics, University of Wisconsin-Madison, Madison, WI 53706, USA \\
$^b$HEP Division, Argonne National Laboratory, 
Argonne, IL 60439, USA
}
}

\begin{document}

\maketitle

\setlength{\parskip}{0.2ex}

\begin{abstract}
The hint of a pseudoscalar toponium state at the Large Hadron Collider (LHC) opens a new avenue for studying a novel class of QCD (quasi-)bound states with comparable formation and decay times. Compared with charmonium and bottomonium, toponium is a quasi-bound state, resembling a hydrogen atom of the strong interaction, although it appears as a broader resonance. We compute the masses and annihilation decay widths of the lowest $S$-wave ($\eta_t$, $\psi_t$) and $P$-wave ($\chi_{t0}$, $\chi_{t1}$) toponium states, and assess their discovery prospects at the High-Luminosity LHC (HL-LHC) and future lepton colliders, such as the $e^+e^-$ stage of the Future Circular Collider (FCC-ee). Detecting the vector $\psi_t$ state at the HL-LHC is hindered by the Landau–Yang theorem and the gluon-dominated production environment of the collider, whereas lepton colliders offer promising sensitivity through both constituent and two-body decays. A more precise measurement of the $\eta_t$ mass—approximately equal to that of $\psi_t$—at the LHC could help determine the optimal $t\bar{t}$ threshold center-of-mass energy for FCC-ee. The $P$-wave states remain challenging to observe at both the HL-LHC and future lepton colliders. We also discuss how toponium measurements can be used to probe top-quark properties and to conduct indirect searches for new physics, including light scalars that couple to the top quark.
\end{abstract}

\thispagestyle{empty}  
\newpage    
\setcounter{page}{1}  

\begingroup
\hypersetup{linkcolor=black,linktocpage}
\tableofcontents
\endgroup

\newpage

\section{Introduction}\label{sec:intro} 

Toponium, the quasi-bound state~\footnote{The binding energy of toponium, or equivalently the inverse of the bound-state formation time, is comparable to the decay width of the constituent top and anti-top quarks. As a result, the two-body system barely completes even a single round of orbital motion before the constituents decay. In this sense, the term ``quasi-bound state" may be more appropriate, although throughout this paper we will use ``toponium'' and ``quasi-bound state'' interchangeably. This behavior is related to the fact that the corresponding pole of the scattering amplitude lies relatively far from the real axis on the second Riemann sheet of the complex energy plane~\cite{Eden:1966dnq}.
} formed by a top quark $t$ and an anti-top quark $\bar{t}$, is the smallest bound state known to date in the Standard Model (SM). Characterized by a Bohr radius of $\sim 0.01\,\mbox{fm}$, the properties of toponia are predominantly captured by perturbative Quantum Chromodynamics (pQCD) in the short-distance regime, hence providing a novel probe for QCD properties including its asymptotic freedom behavior. Moreover, as the threshold production of $t\bar{t}$ at lepton colliders provides an efficient way to measure the properties of the top quark, such as its mass $m_t$ and decay width $\Gamma_t$~\cite{Bigi:1986jk}, the study of the quasi-bound-state effects of toponia is crucial for the complete understanding of the SM.

The quest for toponium at colliders has long been discussed in the literature, both at hadron colliders~\cite{Barger:1987xg,
Fabiano:1993vx,
Kiyo:2008bv,
Hagiwara:2008df,
Kats:2009bv,
Sumino:2010bv,
Kuhn:2013zoa,
Fuks:2021xje,
Aguilar-Saavedra:2024mnm,
Garzelli:2024uhe,
Fuks:2024yjj,
Fuks:2025sxu} 
and lepton colliders~\cite{Fadin:1987wz,
Strassler:1990nw,
Beenakker:1991kh,
Sumino:1992ai,
Murayama:1992mg,
Fujii:1993mk,
Sumino:1994fp,
Harlander:1995dp,
Sumino:1997ve,
Melnikov:1998pr,
Yakovlev:1998ke,
Penin:1998mx,
Nagano:1999nw,
Kuhn:1999hw,
Beneke:1999qg,
Hoang:1999zc,
Hoang:2001mm,
Martinez:2002st,
Eiras:2006xm,
Hoang:2010gu,
Beneke:2010mp,
Penin:2011gg,
Beneke:2013jia,
Beneke:2015lwa,
Bach:2017ggt,
Fu:2024bki,
Defranchis:2025auz} (some more general studies on toponium phenomenology can be found in  Refs.~\cite{Kuhn:1987ty,Yang:2024hnd,
Francener:2025tor}). 
From the theoretical perspective, it is also important to fully understand the QCD potential between the constituent $t$ and $\bar{t}$ to precisely predict the properties of toponia, which has long been an active field of research until now~\cite{Kuhn:1987ty,
Kummer:1994bq,
Fabiano:1994cz,
Kawanai:2013aca,
Akbar:2024brg,
Wang:2024hzd,
Jiang:2024fyw}. Without loss of generality, there are three main methodological categories for the phenomenological predictions near the $t\bar{t}$ threshold: 1. quantum-mechanical-like spectroscopy calculated with the wavefunction projection method (for both production cross sections and decays)~\cite{Barger:1987xg,
Fabiano:1993vx,
Fabiano:1994cz,
Yang:2024hnd,
Wang:2024hzd,
Jiang:2024fyw}, 2. the Green-function method which takes the large-width effect of top quarks into account~\cite{Kiyo:2008bv,
Hagiwara:2008df,
Kats:2009bv,
Fuks:2024yjj,
Strassler:1990nw,
Sumino:1992ai,
Murayama:1992mg,
Fujii:1993mk,
Sumino:1994fp,
Yakovlev:1998ke,
Kuhn:1999hw}, and 3. effective field theories that provide more systematic controls over theoretical uncertainties~\cite{Sumino:2010bv,
Hoang:1999zc,
Beneke:2011mq,
Sumino:1997ve,
Melnikov:1998pr,
Nagano:1999nw,
Beneke:1999qg,
Fuks:2021xje,
Hoang:2001mm,
Hoang:2010gu,
Beneke:2010mp,
Penin:2011gg,
Aguilar-Saavedra:2024mnm,
Bach:2017ggt,
Beneke:2013jia,
Fu:2024bki,
Fu:2025zxb}.
We note here that these methods are not necessarily mutually exclusive and can be combined.
Furthermore, the $t\bar{t}$ ``resonance'' can offer insights into potential new physics, such as that the resonance actually comprises a pseudoscalar state of a mass around $2m_t$~\cite{Djouadi:2024lyv,Lu:2024twj} or that there are exotic short-range interactions ``nailing'' $t$ and $\bar{t}$ together on top of the gluonic force~\cite{Llanes-Estrada:2024phk}.
Besides high energy physics, the resonance production near the $t\bar{t}$ threshold also offers a wonderful playground for the exploration of quantum mechanical properties of particles, including polarizations, spin correlations, entanglement, and the testing of Bell non-locality~\cite{Fadin:1994pj,
Jezabek:1998pj,
Maltoni:2024csn,
Aguilar-Saavedra:2024fig,
Han:2024ugl,
Nason:2025hix}.

Recently, both CMS and ATLAS have reported the cross-section enhancement near
the $t\bar{t}$ production threshold in the pseudoscalar channel at the Large Hadron Collider (LHC) with a significance of over 5\,$\sigma$~\cite{CMS:2025kzt,ATLAS:2026dbe}. This enhancement could receive a sizable contribution from the pseudoscalar toponium state $\eta_t$. However, the Coulomb-enhanced continuum slightly above the threshold may also contribute significantly (see Table 6 of Ref.~\cite{Beneke:2011mq}), given the relatively large bin size of the $t\bar{t}$ invariant mass used in experimental searches. Fixing the top quark Monte Carlo mass $m_t^{\rm MC}=172.5$~GeV and $M_{\eta_t}=343$~GeV (which implies a binding energy of $E_{\rm bind}=M_{\eta_t}-2m_t^{\rm MC}=-2$~GeV) in the Monte Carlo simulation, the production cross section is measured to be $\sigma(pp\to\eta_t)=8.8\substack{+1.2\\-1.4}$~pb by CMS and $9.3\substack{+1.4\\-1.3}$~pb by ATLAS. This opens an avenue for the study of a novel class of QCD (quasi-)bound states with a bound state formation time of $\mathcal{O}[(\alpha_s^2 m_t)^{-1}]$ that is comparable to their decay time of $(2\Gamma_t)^{-1}$. Because of the large top quark decay width, the toponium states behave as relatively broader resonances compared to the charmonium and bottomonium states. Furthermore, the leading discovery channel of toponium states relies on the leading constituent decay channel and not on the suppressed (with branching ratios of less than $10^{-3}$) annihilation decay channels, which is opposite to the charmonium/bottomonium system.
 
Because of their small relative velocity, the binding force between the constituent $t$ and $\bar{t}$ within a toponium system at the production threshold can be well described by the static QCD potential~\cite{Peter:1996ig,Peter:1997me,Schroder:1998vy,Vairo:2007id}, with which one can solve the corresponding non-relativistic Schr\"{o}dinger equation to obtain several properties of the toponium, such as its binding energy (and thus total bound-state mass) and wavefunction. Due to the large top quark decay width $\Gamma_t=1.42$~GeV~\cite{ParticleDataGroup:2024cfk}, the dominant decay mode of a toponium is its intrinsic decay via the decays of individual constituents $t$ and $\bar{t}$. Nevertheless, as we will discuss more in Section~\ref{sec:sensitivity}, some $t\bar{t}$-annihilation decay modes of a toponium can provide additional probes for other SM properties such as the top Yukawa coupling~\cite{Barger:1988zt,Feigenbaum:1990sr}. To study these decays, one can use the projection method~\cite{Kuhn:1980gw,Kuhn:1987ty} to isolate the partial-wave toponium state of interest, which can also be used to study the production of toponia at colliders. However, since the projection method does not take into account the interplay between the continuous $t\bar{t}$ and bound-state toponium production mechanisms, certain quantities (such as the differential cross section $\dd\sigma/\dd m_{t\bar{t}}$, $m_{t\bar{t}}$ being the invariant mass of the $t\bar{t}$ system) must be computed using other methods. For example, the Green function method~\cite{Fadin:1987wz,Strassler:1990nw, Beneke:1999qg, Hoang:2000yr, Hagiwara:2008df, Kats:2009bv, Sumino:2010bv} makes use of the optical theorem to incorporate bound-state effects into the continuous spectrum, although how these two effects should be distinguished is still a topic under active research, as we will explain more in Section~\ref{sec:properties}. Nevertheless, as we show later, these two methods give comparable $pp\to\eta_t$ cross sections around and below the threshold that are also consistent with the CMS~\cite{CMS:2025kzt} and ATLAS~\cite{ATLAS:2026dbe} measurements, which allows the use of the projection method for our purpose of surveying the sensitivity of several toponium states at the current LHC and other future colliders.

As mentioned earlier, since the study of toponia is crucial for the complete understanding of the SM, it is of great importance to survey other possible search channels than the $pp\to\eta_t\to bW^+\bar{b}W^-$ process at the current LHC. For example, in this study, we find that the $\eta_t b\bar{b}$ production can be sensitive at the High-Luminosity LHC (HL-LHC). On the other hand, although it is difficult for $\eta_t$ to be directly produced at lepton colliders due to its $J^{PC}$ quantum numbers, $\psi_t$ turns out to be a fairly promising state to be probed at these facilities, which has been pointed out for long in the literature (see for example Refs.~\cite{Sumino:1992ai,Fujii:1993mk,Nagano:1999nw,Beneke:1999qg}). On the contrary, the direct production of $gg\to\psi_t$ is prohibited by the Landau-Yang theorem~\cite{Landau:1948kw,Yang:1950rg}, and thus it is challenging to discover $\psi_t$ at hadron colliders.
In this study, we especially discuss the $e^+e^-$ stage of the Future Circular Collider (FCC-ee)~\cite{Schwienhorst:2022yqu, FCC:2025lpp, FCC:2025uan}, the preferred option for CERN suggested by the European Strategy Group~\cite{FCC-ESPP2026}, which proposes to run around the $t\bar{t}$ threshold with an integrated luminosity of $\mathcal{O}(1)~{\rm ab}^{-1}$. For $\psi_t$, in addition to the individual $t$ and $\bar{t}$ decays, Ref.~\cite{Fu:2024bki} shows that its leading $t\bar{t}$-annihilation decay to $b\bar{b}$ can also be probed via some specially designed observable that captures the interference with the $\gamma$- and $Z$-mediated processes caused by the $\psi_t$ mass pole. In this study, we independently calculate the production cross section including the interference effects and identify a sign mistake in Ref.~\cite{Fu:2025zxb} (their earlier work Ref.~\cite{Fu:2024bki} and the public model file are both correct), which leads to a larger cross section in their result compared to the correct numbers presented in this study.
Moreover, one can also perform sensitive studies on other promising channels, such as processes $e^+e^- \to \psi_t \to W^+W^-, \gamma Z, \text{ and } \gamma H$.

Finally, in addition to probing the SM, one can also constrain certain beyond-SM (BSM) physics with toponium measurements, such as new scalar fields that either affect the top Yukawa coupling of the 125-GeV Higgs boson or contribute additional Yukawa interactions to the static toponium QCD potential. In our study, we use the $\mathbb{Z}_2$-symmetric real-singlet extension to the SM (SSM) as an example and show how the $\eta_t$ measurement performed in Ref.~\cite{CMS:2025kzt} can be used to constrain its parameter space, while one can certainly generalize this idea to other BSM models.

Our paper is organized as follows. In Section~\ref{sec:properties}, we review the theoretical properties of toponia, including their masses, decays, and production at colliders, with the details of the $t\bar{t}$-annihilation decay formulas and the results of $P$-wave Green function calculation summarized in Appendices~\ref{subapp:toponium_decay_width} and \ref{subapp:GF:diff}, respectively. In Section~\ref{sec:sensitivity}, we discuss the discovery sensitivity of the four toponium states $\eta_t$, $\psi_t$, $\chi_{t0}$, and $\chi_{t1}$ at the current LHC and future HL-LHC and FCC-ee. In Section~\ref{sec:SSM}, we perform a parameter space study on the SSM using the $\eta_t$ measurement conducted by CMS as an example of probing BSM physics using toponium. Finally, we conclude our study in Section~\ref{sec:conclusions}.

\section{Properties of toponium states}\label{sec:properties}

In this section, we summarize some basic properties of the toponium states, including their mass spectrum, decays, and production at colliders.

\subsection{Mass spectrum}\label{subsec:spectrum}

The binding energy and wavefunction of a heavy quarkonium near the threshold can be obtained by solving the non-relativistic Schr\"{o}dinger equation given the small velocities of the constituent quarks. For a color-singlet toponium system in the short-distance regime, the associated static potential described by pQCD up to two-loop is given by~\cite{Peter:1996ig,Peter:1997me,Schroder:1998vy,Vairo:2007id,ParticleDataGroup:2024cfk,Sumino:2010bv}
\begin{align}
    V_{\rm pQCD}(r)=&~-\frac{C_F\,\alpha_s(\mu_R)}{r}\left\{1+\frac{\alpha_s(\mu_R)}{4\pi}[a_1+2\gamma_E\beta_0]\right. \label{eqn:potential}\\  &~\left.\hspace{3.0cm}+\left(\frac{\alpha_s(\mu_R)}{4\pi}\right)^2 \left[a_2+\left(\frac{\pi^2}{3}+4\gamma_E^2\right)\beta_0^2+2\gamma_E(2a_1\beta_0+\beta_1)\right]\right\} ~, \nonumber
\end{align}
where $\gamma_E\approx 0.577$ is the Euler-Mascheroni constant and  $\alpha_s(\mu_R)$ is the strong coupling at the renormalization scale $\mu_R$ as given by
\begin{align}
    \alpha_s(\mu_R)=\frac{4\pi}{\beta_0\log(\mu_R^2/\Lambda^2)}\left\{1-\frac{\beta_1}{\beta_0^2}\frac{\log\left[\log(\mu_R^2/\Lambda^2)\right]}{\log(\mu_R^2/\Lambda^2)}\right\} ~.
\end{align}
A convenient scale choice of $\mu_R$ is $\sim 1/r$, which we will stick to in this study and check the related variation later. The color $SU(3)$ group constants are defined through $C_F=4/3$, $C_A=3$, $T_F=1/2$, $n_f=5$, $\beta_0=11C_A/3-4T_F n_f/3$, $\beta_1=34C_A^2/3-20C_AT_F n_f/3-4C_F T_F n_f$, and
\begin{align}
    a_1=&~\frac{31}{9}C_A-\frac{20}{9}T_Fn_f ~,\\
    a_2=&~\left[\frac{4343}{162}+4\pi^2-\frac{\pi^4}{4}+\frac{22}{3}\zeta(3)\right]C_A^2-\left[\frac{1798}{81}+\frac{56}{3}\zeta(3)\right]C_AT_Fn_f\nonumber\\
    &~-\left[\frac{55}{3}-16\zeta(3)\right]C_F T_F n_f+\left[\frac{20}{9}T_F n_f\right]^2 ~.
\end{align}
In the long-distance regime, one can replace the potential in Eq.~\eqref{eqn:potential} with the Cornell potential 
\begin{equation}
    V_{\rm Cornell} = -\frac{C_F\,\alpha_s^{\rm Cornell}}{r}+b\,r ~,    
\end{equation}
starting from the transition point $r=r_0=0.1$~fm, where $\alpha_s^{\rm Cornell}=0.211$ and $b=0.212$~GeV$^2$ are taken to be fixed values evaluated at the transition point (see Ref.~\cite{Koma:2007jq} for the results from Lattice QCD calculations). Overall, one has the complete static potential
\begin{equation}\label{eq:V_static}
    V_{\rm static}(r) = \begin{cases}
        ~V_{\rm pQCD}(r) &,\quad  r\leq r_0 \\[1ex]
        ~V_{\rm Cornell}(r) &,\quad r > r_0
    \end{cases} ~.
\end{equation}
Note that the Bohr radius of the toponium states $a_t$ can be estimated using the Coulomb potential as $a_t=(C_F \, \alpha_s \, m_t/2)^{-1}\sim (18~{\rm GeV})^{-1}$. One then has $\alpha_s=\alpha_s(a_t^{-1})\approx\alpha_s(18~{\rm GeV})\approx 0.157$ and $a_t\approx0.011$~fm~\cite{ParticleDataGroup:2024cfk}, which is one order of magnitude smaller than the transition radius $r_0$. This also means that the properties of toponia are dominated by the short-distance pQCD potential.

To calculate the mass splitting among the different toponium states with the same orbital angular momentum, one can use the spin-dependent potential~\cite{Barnes:2005pb}
\begin{align}
    V_{\text{spin-dep}}(r)=\frac{8\,\alpha_s(\mu_R)}{9\,m_t^2\,r^2}\delta(r)\vv{\mbox{S}}_t\cdot\vv{\mbox{S}}_{\bar{t}}+\frac{1}{m_t^2}\left[\left(\frac{2\alpha_s(\mu_R)}{r^3}-\frac{b}{2r}\right)\vv{\mbox{L}}\cdot\vv{\mbox{S}}+\frac{4\alpha_s(\mu_R)}{r^3}\mbox{T}\right] ~.\label{eqn:splitting-potential}
\end{align}
Here, $\vv{\mbox{S}}_t$ and $\vv{\mbox{S}}_{\bar{t}}$ are the spin operators of the top and anti-top quarks, respectively.  
$\vv{\mbox{S}}=\vv{\mbox{S}}_t+\vv{\mbox{S}}_{\bar{t}}$ is the total spin angular momentum operator, $\vv{\mbox{L}}$ is the orbital angular momentum operator, and $\vv{\mbox{J}}=\vv{\mbox{S}}+\vv{\mbox{L}}$ is the total angular momentum operator. We denote $S,L,J$ as the eigenvalues of the corresponding angular momentum operators. The tensor operator $\mbox{T}$ has non-zero eigenvalues only between the $L>0$ spin-triplet states as 
\begin{equation}
    \langle^3L_J\vert \,\mbox{T}\, \vert^3L_J\rangle = \begin{cases}
        ~-\frac{L}{6(2L+3)} &,\quad J=L+1 \\[1ex]
        ~+\frac{1}{6} &,\quad J=L \\[1ex]
        ~-\frac{L+1}{6(2L-1)} &,\quad J=L-1
    \end{cases} ~.
\end{equation}

\begin{table}[bht!]
    \centering
    \renewcommand{\arraystretch}{1.5}
    \begin{tabular}{c|c|c|c|c}
    \hline\hline
        \multirow{2}{*}{toponium} & \multicolumn{2}{c|}{binding energy [GeV]} & \multicolumn{2}{c}{$R_S(0)$~[$a_t^{-3/2}$],\quad  $R_P'(0)$~[$a_t^{-5/2}$]} \\ \cline{2-5}
        & $m_t=172.4\pm{0.7}~{\rm GeV}$ & $\mu_R=(1.0\substack{+1.0\\-0.5})/r$ & $m_t$ & $\mu_R$  \\ \hline\hline
        $S$-wave & $-2.725_{-0.005}^{+0.005}$ & $-2.725_{-1.020}^{+0.654}$ & $1.611_{-0.001}^{+0.001}$ & $1.611_{-0.205}^{+0.235}$ \\
        $P$-wave& $-1.375_{-0.003}^{+0.003}$ & $-1.375_{-0.813}^{+0.430}$ & $0.2346_{-0.0003}^{+0.0003}$ & $0.2346_{-0.0444}^{+0.0208}$ \\ \hline\hline
    \end{tabular}
    \renewcommand{\arraystretch}{1}
    \caption{The binding energies and (derivatives of) the radial wavefunctions at the origin for the lowest $S$-wave and $P$-wave toponium states obtained by solving the Schr\"odinger equation with the full $V_{\rm static}(r)$ given in Eq.~\eqref{eq:V_static}. The Bohn radius is $a_t=[C_F \, \alpha_s(a_t^{-1}) \, m_t/2]^{-1}$. The uncertainties come either from the experimental uncertainty in $m_t$ or the variation of the renormalization scale $\mu_R$ from $0.5/r$ to $2/r$.  
    }
    \label{tab:mass_wf_fullpotenial_Coulomb}
\end{table} 

\begin{figure}[th!]
    \centering
    \includegraphics[width=0.7\linewidth]{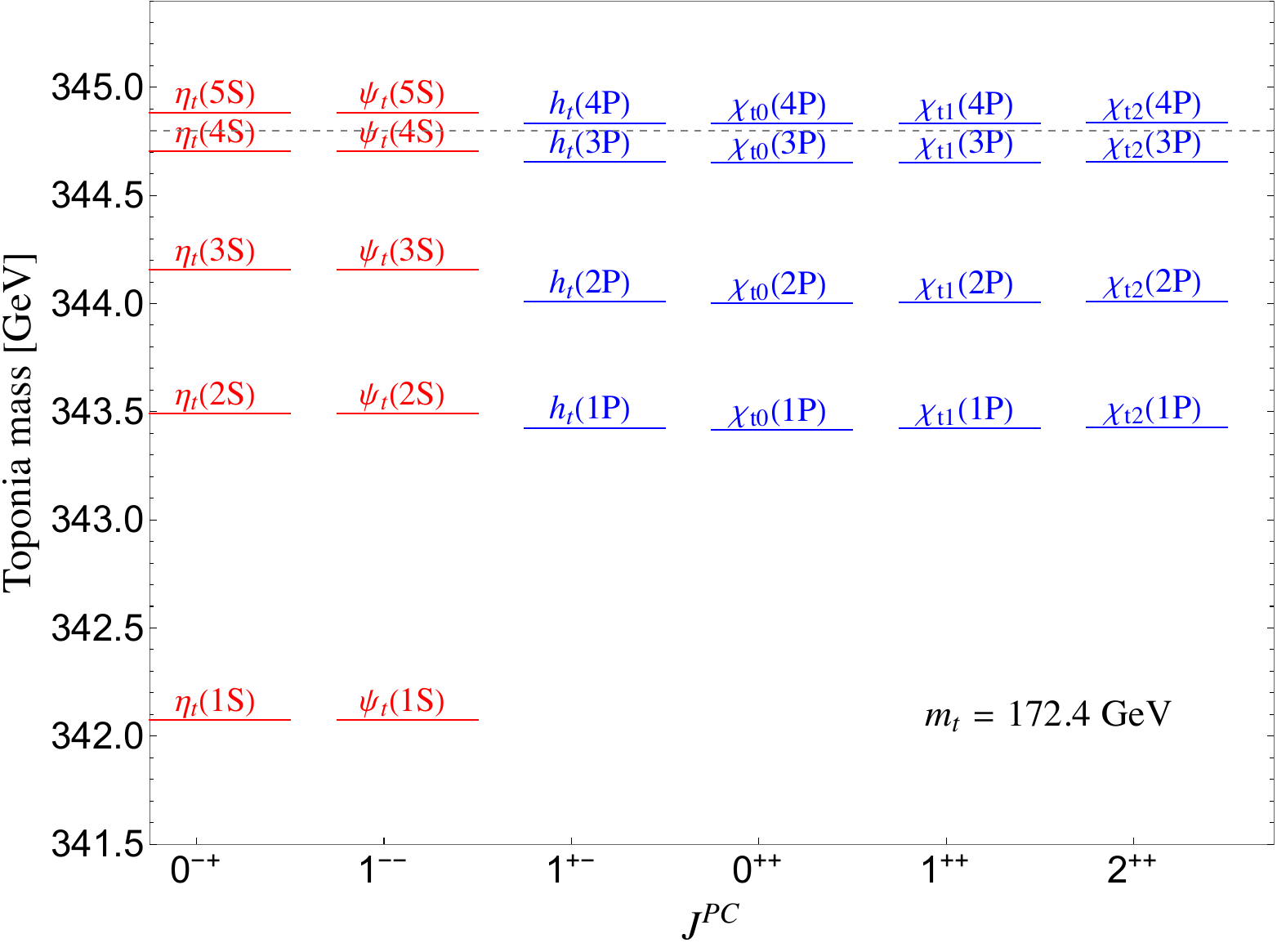}
    \caption{The mass spectrum of the lowest $S$- and $P$-wave toponium states, categorized by their principal quantum number $n$ and $n-1$ and $J^{\rm PC}$ quantum numbers. In this plot, we fix $m_t=172.4$~GeV, and the gray dashed line denotes the threshold mass $2m_t=344.8$~GeV, above and below which $E_{\rm bind}>0$ and $E_{\rm bind}<0$, respectively. The renormalization scale for $\alpha_s(\mu_R)$ in the static potential is chosen to be $\mu_R = 1/r$. There is an uncertainty of around $\delta M\sim 2$~GeV for all toponium masses resulting from the top quark mass and renormalization scale uncertainties. The mass difference between $\eta_t(n\mbox{S})$ and $\psi_t(n\mbox{S})$ is negligible and that among the $P$-wave states with the same principle quantum number $n-1$ is around $5~{\rm MeV}$. On the other hand, the mass splitting between $S$- and $P$-wave states with the same principle quantum number $n$ is sizable, \eg, the mass difference between $\eta_t(2S)$ and $h_t(1P)$ is $66~{\rm MeV}$.
    }
    \label{fig:toponia_spectrum}
\end{figure}

By solving the Schr\"odinger equation with the full static potential given in Eq.~\eqref{eq:V_static}, one can obtain the binding energy $E_{\rm bind}$ (which can be either positive or negative) and wavefunction of a given toponium state. The bound-state mass of a toponium is then given by $M = 2 \, m_t + E_{\rm bind}$, where the top quark mass is taken to be the measured pole mass $m_t =m_t^{\rm pole} =172.4\pm{0.7}~{\rm GeV}$~\cite{ParticleDataGroup:2024cfk}. 
The uncertainty in $m_t$ will introduce an uncertainty in the toponium mass of around $\delta M \approx 1.4$~GeV. Moreover, since there is no definitive way to fix $\mu_R$, the range of this choice will lead to an additional uncertainty of $\mathcal{O}(1\,\mbox{GeV})$ in $\delta M$ through $\alpha_s(\mu_R)$, which can be seen in Table~\ref{tab:mass_wf_fullpotenial_Coulomb}, in which we investigated the binding energies and the values at the origin of the wavefunctions $R_S(r)$ [the derivatives of wavefunctions $R_P^\prime(r)$] for the $S$-wave ($P$-wave) states, respectively, by varying $\mu_R$ from $0.5/r$ to $2/r$. Overall, the uncertainty in the predicted toponium mass is $\delta M \simeq 2$~GeV. Using the obtained wavefunctions and Eq.~\eqref{eqn:splitting-potential}, we calculate the mass splittings among the states with the same $L$ and show the spectrum of the first few $S$- and $P$-wave states in Figure~\ref{fig:toponia_spectrum}, where we fix the top quark mass $m_t$ at its central value $172.4$~GeV and the renormalization scale $\mu_R = 1/r$. Note that all toponia have an uncertainty of around $\delta M\sim 2$~GeV coming from the uncertainties in $m_t$ and $\mu_R$ as mentioned previously, while the relative mass differences are all much smaller and thus negligible. 

Note that because $\alpha_s(a_t^{-1})/\pi\approx0.05\ll1$ for a typical toponium system, it can be well described by pQCD through the leading-order Coulomb potential $V_{\rm Coul}(r)=-C_F\alpha_s^{\rm Coul}/r$, where $\alpha_s^{\rm Coul}$ is a fixed number whose value we will determine later, with both the loop correction and linear potential neglected. Likewise, the other $S,L,J,T$-dependent terms are subdominant compared to $V_{\rm Coul}$ and will only produce small splittings among the masses of the different toponium states, which we label using the term symbol $^{2S+1}L_J$. As a result, as far as the wavefunctions of the different toponium states are concerned, taking $V\approx V_{\rm Coul}$ is in general a good approximation. 
To demonstrate this, we fix $\alpha_s^{\rm Coul}$ by matching the derived binding energy of the $S$-wave states to that calculated from using the full $V_{\rm static}(r)$ given in Eq.~\eqref{eq:V_static}. This matching leads to $\alpha_s^{\text{Coul}}=0.189$. We then compare the values of the radial wavefunctions of the two lowest $S$-wave states $\eta_t(^1S_0),\psi_t(^3S_1)$ [$R_S(r)$] and the derivatives of those of the two lowest $P$-wave states $\chi_{t0}(^3P_0),\chi_{t1}(^3P_1)$ [$R_P^\prime(r)$] at the origin $r=0$ using both $V_{\rm Coul}(r)$ and the full $V_{\rm static}(r)$. We list the latter in Table~\ref{tab:mass_wf_fullpotenial_Coulomb}, while the former are given by
\begin{align}
    R_S(0)=2\left(\frac{4}{3}\frac{\alpha_s^{\text{Coul}} m_t}{2}\right)^{\frac{3}{2}}= 2(a_t^{\text{Coul}})^{-\frac{3}{2}},~~R_P'(0)=\frac{1}{\sqrt{24}}\left(\frac{4}{3}\frac{\alpha_s^{\text{Coul}} m_t}{2}\right)^{\frac{5}{2}}\approx0.2041(a_t^{\text{Coul}})^{-\frac{5}{2}} ~,\label{eqn:waveFun_Coulomb}
\end{align}
with the binding energies of the $P$-wave states being $-0.6812~{\rm GeV}$. One can see that the two methods indeed give close though not identical results. This discrepancy can be explained by the following reasons. First, when $r\to 0$, the full potential $V_{\rm static}$
will be suppressed through the running of $\alpha_s(\mu_R\sim 1/r)$ due to the asymptotic freedom of QCD, while $V_{\rm Coul}$ remains parametrized by the fixed $\alpha_s^{\rm Coul}$. This difference between the behavior of the two potentials at short distance will then influence (the derivatives of) the wavefunctions at the origin. Second, the $P$-wave wavefunctions are distributed considerably about $r\sim \mathcal{O}(1)a_t$, which causes them to receive more corrections from the long-distance contribution of the static potential compared to the $S$-wave states. We further scan $\kappa\equiv\mu_R\times r$ and find that the larger $\kappa$, the more similar the $S$-wave wavefunctions derived with $V_{\rm static}$ and $V_{\rm Coul}$ will be. 
This is because $\alpha_s(\mu_R=\kappa/r)$ will decrease along with shorter $r$, making the system more similar to the case described only by $V_{\rm Coul}$, which after matching $E_{\rm bind}=-1.88$~GeV is parametrized by $\alpha_s=\alpha_s(a_t^{-1})=0.157$.
In the remainder of the study, we will only use $V_{\rm Coul}$ with $\alpha_s^{\rm Coul}=0.189$ to calculate the wavefunctions for simplification, which we will discuss more in Section~\ref{subsec:eta_t}.

\subsection{Decays}\label{subsec:decay}

The dominant decay mode of toponia is the intrinsic decays of the constituent top and anti-top quarks with the decay width $\Gamma\approx\Gamma_t+\Gamma_{\bar{t}}=2\Gamma_t=2.84~{\rm GeV}$~\cite{ParticleDataGroup:2024cfk}. On top of that, one can also consider the $t\bar{t}$-annihilation decay modes, which have been considered in Refs.~\cite{Kuhn:1980gw,Kuhn:1987ty} and calculated using the ``projection method'', which works as follows: Consider a generic $Q\bar{Q}\to f$ annihilation process with the matrix element
\begin{align}
    \mathcal{M}(Q\bar{Q}\to f)=\bar{v}^s\left(\frac{p}{2}-q\right)\mathcal{O}(p,q)u^{s'}\left(\frac{p}{2}+q\right) ~,
\end{align}
from which we further define $\mathcal{O}(0)\equiv\mathcal{O}(p, q=0)$ and $\mathcal{O}^\alpha(0)\equiv\frac{\partial}{\partial q_\alpha}\mathcal{O}(p, q)\vert_{q=0}$. Consequently, one can write down the bound-state version of the matrix element $\mathcal{M}[B(^{2S+1}L_J)\to f]\equiv \mathcal{A}(^{2S+1}L_J)$ in the non-relativistic limit, where $B(^{2S+1}L_J)$ is the $Q\bar{Q}$ bound state characterized by the quantum numbers $S,L,J$. For the $S$-wave states, one has
\begin{align}
    &\mathcal{A}(^1S_0)=\frac{1}{\sqrt{3}}\frac{1}{\sqrt{M_{B}}}\frac{1}{2}\frac{1}{\sqrt{4\pi}}R_S(0)\Tr[\mathcal{O}(0)(M_{B}+\cancel{p})(-\gamma^5)] ~,\label{eqn:projection_s0l0j0}\\
    &\mathcal{A}(^3S_1)=\frac{1}{\sqrt{3}}\frac{1}{\sqrt{M_{B}}}\frac{1}{2}\frac{1}{\sqrt{4\pi}}R_S(0)\Tr[\mathcal{O}(0)(M_{B}+\cancel{p})(-\gamma^\mu)]\varepsilon_\mu(p)\label{eqn:projection_s0l0j1} ~,
\end{align}
where $\varepsilon_\mu(p)$ is the polarization vector of the $^3S_1$ state, which is the $\psi_t$ state when it comes to toponium. Note that the trace is also taken over the color indices, while the prefactor $\frac{1}{\sqrt{3}}$ arises from the projection onto the color-singlet component of the $Q\bar{Q}$ bound state. As for the $P$-wave states, one has
\begin{align}
    &\hspace{-2mm}\mathcal{A}(^1P_1)=\frac{1}{\sqrt{3}}\frac{i}{\sqrt{M_B}}\frac{1}{2}\sqrt{\frac{3}{4\pi}}R_P^{'}(0)\Tr[\big[\mathcal{O}^\mu(0)-\frac{2}{M_B}\mathcal{O}(0)\gamma^\mu\big](M_{B}+\cancel{p})\gamma^5]\varepsilon_\mu(p) ~,\label{eqn:projection_s0l1j1}\\
    &\hspace{-2mm}\mathcal{A}(^3P_0)=\frac{1}{\sqrt{3}}\frac{i}{\sqrt{M_B}}\sqrt{\frac{1}{4\pi}}R_P^{'}(0)\Tr[\frac{1}{2}\mathcal{O}^\mu(0)(\gamma_\mu+\frac{p_\mu}{M_B})(\cancel{p}-M_{B})+3\mathcal{O}(0)] ~,\label{eqn:projection_s1l1j0}\\
    &\hspace{-2mm}\mathcal{A}(^3P_1)=\frac{1}{\sqrt{3}}\frac{1}{\sqrt{M_B}}\sqrt{\frac{3}{8\pi}}R_P^{'}(0)\epsilon_{\alpha\beta\gamma\delta}\frac{p^\gamma}{M_B}\Tr[\frac{1}{2}\mathcal{O}^\alpha(0)\gamma^\beta(\cancel{p}-M_{B})-\mathcal{O}(0)\gamma^\alpha\gamma^\beta \frac{\cancel{p}}{M_B}]\varepsilon^\delta(p) ~,\label{eqn:projection_s1l1j1}\\
    &\hspace{-2mm}\mathcal{A}(^3P_2)=\frac{1}{\sqrt{3}}\frac{i}{\sqrt{M_B}}\frac{1}{2}\sqrt{\frac{3}{4\pi}}R_P^{'}(0)\Tr[\mathcal{O}^\alpha(0)\gamma^\beta(\cancel{p}-M_{B})]\varepsilon_{\alpha\beta}(p)\label{eqn:projection_s1l1j2} ~.
\end{align}
Later in this study, we will focus on the phenomenology of $\eta_t$, $\psi_t$, $\chi_{t0}$, and $\chi_{t1}$, the branching ratios of whose annihilation decay modes are summarized in Table~\ref{tab:decay_width_projection_value} with the detailed corresponding formulas listed in Appendix~\ref{subapp:toponium_decay_width} (see also Refs.~\cite{Barger:1987xg, Fu:2025zxb}). Here we take the massless limits for the fermions, and thus the fermionic decay modes of $\eta_t$ and $\chi_{t0}$, whose couplings to fermion pairs are proportional to the fermions masses, become negligible, with the exception of the $\eta_t\to b\bar{b}$ decay, for which we retain the non-zero bottom quark mass during the calculation. Note that all annihilation decay modes have partial widths that are significantly smaller than the intrinsic decay width $\Gamma$, and some of them are prohibited by $J^{PC}$ conservation and thus have exactly zero partial widths. We will discuss the sensitivity of toponium states in some of these channels at the LHC and future colliders in Section~\ref{sec:sensitivity}.
\begin{table}[ht!]
    \centering
    \renewcommand{\arraystretch}{1.25}
    \begin{tabular}{c|c|c|c|c}\hline\hline
        decay mode & \multirow{2}{*}{$\eta_t~(0^{-+})$} & \multirow{2}{*}{$\psi_t~(1^{--})$} & \multirow{2}{*}{$\chi_{t0}~(0^{++})$} & \multirow{2}{*}{$\chi_{t1}~(1^{++})$} \\ 
        (branching ratio) & & & & \\ \hline\hline
        $W^+W^-$ & $2.42\times10^{-4}$ & $9.96\times10^{-4}$ & $1.28\times10^{-6}$ & $1.45\times10^{-6}$\\
        $ZZ$ & $1.22\times10^{-5}$ & $9.29\times10^{-6}$ & $1.70\times10^{-7}$ & $1.47\times10^{-8}$\\
        $ZH$ & $1.02\times10^{-3}$ & $1.38\times10^{-5}$ & 0 & $9.84\times10^{-7}$ \\
        $H H$ & 0 & 0 & $4.34\times10^{-6}$ & 0 \\
        $\gamma\gamma$ & $1.78\times10^{-5}$ & 0 & $2.65\times10^{-8}$ & 0 \\
        $\gamma H$ & 0 & $1.14\times10^{-4}$ & 0 & 0 \\
        $\gamma Z$ & $3.94\times10^{-6}$ & $1.32\times10^{-4}$ & $6.47\times10^{-9}$ & $1.15\times10^{-10}$ \\ \hline
        $\nu \bar{\nu}$ & - & $1.30\times10^{-6}$ & - & $8.62\times10^{-9}$ \\
        $\ell^+\ell^-$ & - & $1.45\times10^{-5}$ & - & $4.34\times10^{-9}$ \\
        $u\bar {u},c\bar {c}$ & - & $2.47\times10^{-5}$ & - & $1.49\times10^{-8}$\\
        $d\bar {d},s\bar {s}$ & - & $1.14\times10^{-5}$ & - & $1.91\times10^{-8}$\\
        $b\bar{b}$ & $7.36\times10^{-8}$ & $8.54\times10^{-4}$ & - & $3.34\times10^{-8}$ \\ \hline
        $gg$ & $4.58\times10^{-3}$ & 0 & $6.81\times10^{-6}$ & 0\\
        $ggg$ & 0 & $2.77\times10^{-5}$ & 0 & 0 \\  \hline\hline
    \end{tabular}
    \renewcommand{\arraystretch}{1}
    \caption{The branching ratios of the $t\bar{t}$-annihilation decay modes of $\eta_t$, $\psi_t$, $\chi_{t0}$, and $\chi_{t1}$. Some decay modes are prohibited by $J^{PC}$ conservation and thus have exactly zero partial widths. Here we ignore the fermion masses, and thus the fermionic decay widths of $\eta_t$ and $\chi_{t0}$ are negligible, denoted as ``-'', except for $\eta_t\to b\bar{b}$, for which we explicitly include the non-zero bottom quark mass in the calculation. For the results with non-zero fermion masses, we refer to Refs.~\cite{Barger:1987xg,Fu:2025zxb}.
    }
    \label{tab:decay_width_projection_value}
\end{table}

\subsection{Production}\label{subsec:prod}

To calculate the production cross sections of toponia, one can use the projection method by modifying the kinematics accordingly and then applying the same formulas as in Eqs.~\eqref{eqn:projection_s0l0j0}-\eqref{eqn:projection_s1l1j2}. For example, the gluon fusion production cross section of an exactly on-shell $\eta_t$ at the parton level is given by
\begin{align}
    \hat{\sigma}(gg\to \eta_{t})_{\rm on-shell}=\frac{\pi^2\alpha_s^2}{3M\hat{s}}|R_S(0)|^2\delta(\hat{s}-M^2) ~,
\end{align}
where $\sqrt{\hat{s}}$ is the parton-level center-of-mass energy. Alternatively, one can also include the finite-width effect by using the Breit-Wigner (B-W) formula, which gives
\begin{equation}
    \hat{\sigma}(gg\to \eta_{t})_{\rm B-W} = \frac{\pi\alpha_s^2}{3M^2\Gamma\hat{s}}|R_S(0)|^2\frac{\Gamma^2/4}{(\sqrt{\hat{s}}-M)^2+\Gamma^2/4} ~.
\end{equation}
In general, the projection method is an appropriate way to compute processes involving bound states, as long as the magnitude of the relative momentum between the constituent quarks $\vert\vec{q}\vert$ is much smaller than the bound-state mass $M$, which is usually the case in reality for the production of heavy quarkonia near the threshold~\cite{Bai:2024ura}. For simplicity, we will use this method to estimate the production cross sections in the sensitivity study in Section~\ref{sec:sensitivity}.

Nevertheless, the projection method ignores the interplay between the continuous $t\bar{t}$ and bound-state toponium production mechanisms, which is important when it comes to scrutinizing the differential cross sections constructed from measuring their identical decay products. To incorporate this interplay into the calculations, one can turn to the ``Green function method''~\cite{Fadin:1987wz, Fadin:1990wx, Beneke:1999qg, Hoang:2000yr, Hagiwara:2008df, Kats:2009bv, Sumino:2010bv}, which is based on the optical theorem and accounts for the enhancement near the threshold by multiplying the continuous production cross section by the ratio of the imaginary part of the bound-state Green function to that of the free Green function. For instance, the gluon fusion production cross section of $bW^+\bar{b}W^-$ through the $S$-wave color-singlet toponium states can be calculated via~\cite{Fadin:1990wx, Hoang:2000yr,Sumino:2010bv}~\footnote{The color-octet channel is subleading around the threshold~\cite{Sumino:2010bv}.}
\begin{align}
    \hat{\sigma}(gg\to t\bar{t}\to bW^+\bar{b}W^-)_{S}=\hat{\sigma}(gg\to t\bar{t} \to bW^+\bar{b}W^-)_{S,\text{tree}}\,\frac{\Im[G(r,E)]_{r=0}}{\Im[G_0(r,E)]_{r=0}} ~,
\end{align}
where $\sigma_{S,{\rm tree}}$ is the Born-level continuous production cross section through on-shell $t\bar{t}$, and $G_0$ and $G$ denote the free and bound-state Green functions of the $t\bar{t}$ system, respectively. The bound-state Green function $G$ is solved from the following Schr\"{o}dinger equation,\footnote{
One can also apply the Green function method in the momentum space as studied in Refs.~\cite{Fuks:2024yjj,Fuks:2025sxu}, which is useful for Monte Carlo simulations of kinematic observables.
} 
\begin{align}\label{eq:Green}
    \left[\frac{\nabla^2}{m_t}-V(r)+z\right]G(\vec{r},z)=\delta^3(\vec{r}) ,\quad z=E+i\Gamma_t ~,
\end{align}
where we approximate the QCD potential $V(r)\approx V_{\rm Coul}(r) = -C_F\alpha_s^{\rm Coul}/r$. The analytic solution to Eq.~\eqref{eq:Green} is given by~\cite{Hostler:1964}
\begin{align}\label{eq:green_analytic}
    G(r,E)=\frac{m_t^2}{4\pi}a\,\Gamma\left(-\frac{i a}{2\sqrt{z'}}\right)e^{i\sqrt{z'}\rho}\,U\left(1-\frac{ia}{2\sqrt{z'}},2,-2i\sqrt{z'}\rho\right) ~,
\end{align}
where we use the dimensionless quantities $\rho=m_t r$, $z'=z/m_t$, and $a=C_F\alpha_s^{\rm Coul}$. Here, $\Gamma(x)$ is the gamma function and $U(\mathfrak{a},\mathfrak{b},x)$ is the Tricomi confluent hypergeometric function. When $a\to 0$, one retrieves the free Green function $G_0$,
\begin{align}
    G_0(r,E)=-\frac{m_t^2}{4\pi}\frac{e^{i\sqrt{z'}\rho}}{\rho} ~.
\end{align}
When $r\to 0$, $G$ has an asymptotic behavior,
\begin{align}
    \lim_{r\to0}G(r,E) \to \frac{m_t^2}{4\pi} &\bigg[-\frac{1}{\rho} +a \log \rho -2ia\pi +a \log(-2i\sqrt{z'}) \nonumber\\
    &\quad+\psi^{(0)}\left(1-\frac{ia}{2\sqrt{z'}}\right) +2\gamma_E a-a -i\sqrt{z'} \bigg] ~,
\end{align}
where $\psi^{(n)}(x)$ is the $n$-th derivative of the digamma function. Consequently, one can define the $S$-wave Sommerfeld enhancement factor as
\begin{align}
    \mathcal{S}_S(E)\equiv\frac{\Im[G(r,E)]_{r=0}}{\Im[G_0(r,E)]_{r=0}}=\frac{\Im\left[-a\log\left(-2i\sqrt{z'}\right)-a\,\psi^{(0)}\left(1-i\frac{a}{2\sqrt{z'}}\right)+i\sqrt{z'}\right]}{\Im\left[i\sqrt{z'}\right]} ~.\label{eqn:SommerfeldEnhancement}
\end{align}
It is worth noting that the denominator in Eq.~\eqref{eqn:SommerfeldEnhancement} will in general cancel the kinematic factor $\sqrt{1-4m_t^2/\hat{s}}$ in the cross section formula when $\Gamma_t\to0$, allowing us to extend the distribution of differential cross section $\dd\sigma/\dd m_{t\bar{t}}$ below the threshold of $2m_t$ without inducing an imaginary number, as shown in Ref.~\cite{Kats:2009bv}.

\begin{figure}[th!]
    \centering
    \includegraphics[width=0.7\linewidth]{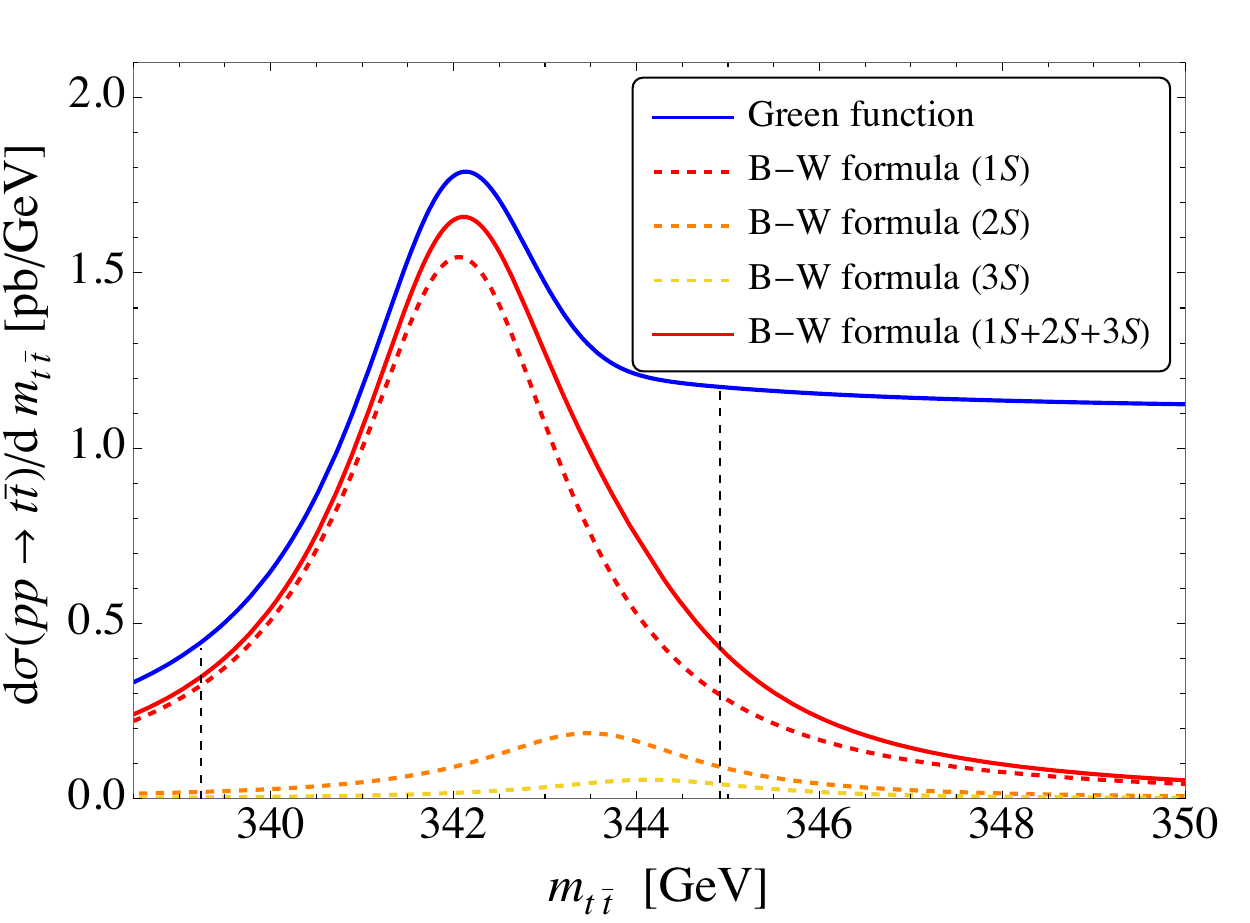}
    \caption{The $\dd\sigma(pp\to\eta_t)/\dd m_{t\bar{t}}$ distributions at the 13-TeV LHC calculated using the leading-order Green function method (blue) and the B-W formula with the first three $\eta_t (nS)$ states superposed (red). The two dashed vertical lines represent $m_{t\bar{t}}=M_{\eta_t}\pm\Gamma$ with $M_{\eta_t}=342.08~{\rm GeV}$ and $\Gamma=2.84~{\rm GeV}$, respectively, between which we identify the $t\bar{t}$ system to be on ``resonance''. We also show the individual $nS$ resonances with $n=1$ (red), $n=2$ (orange), and $n=3$ (yellow) in dashed curves. }
    \label{fig:xs_pp_etat_GreenFunc}
\end{figure}

Here we demonstrate the consistency between the Green function and projection methods regarding the production cross section $\sigma(pp\to\eta_t)$. For the Green function method, we take $V_{\rm Coul}(r)$ with $\alpha_s^{\rm Coul}=0.189$ as obtained from matching the $S$-wave binding energy with $\mu_R = 1/r$ in Section~\ref{subsec:spectrum}, while for the projection method, we use the same setup as in Section~\ref{subsec:decay} when studying the annihilation decays and adopt the B-W formula. We compute the differential cross section $\dd\sigma(pp\to\eta_t)/\dd m_{t\bar{t}}$ with both methods and show the results in Figure~\ref{fig:xs_pp_etat_GreenFunc}. One can see that the two methods provide comparable values around and below the threshold. On the other hand, the Green function method predicts an additional continuum spectrum above the threshold.
Next, we define the ``resonance'' region in the $t\bar{t}$ invariant mass spectrum as $m_{t\bar{t}}\in[M_{\eta_t}-\Gamma,M_{\eta_t}+\Gamma]=[342.08-2.84,342.08+2.84]$~GeV and integrate the differential cross section over this interval to obtain the proton-level production cross section $\sigma(pp\to\eta_t)_{\rm Green}=7.08~{\rm pb}$, which is compatible with the experimental result of $\sigma(pp\to\eta_t)=8.8\substack{+1.2\\-1.4}~{\rm pb}$ reported in Ref.~\cite{CMS:2025kzt} within 2\,$\sigma$. On the other hand, the projection method gives the cross section of $\sigma(pp\to\eta_t)_{\rm Projection}=6.89~{\rm pb}$, which is also compatible with both the Green function and the experimental results. We note here that the experimental measurements reported at the LHC consist of both the quasi-bound state and above-threshold-continuum contributions. Also, the $K$-factor is very close to one and thus does not influence the results here much, as mentioned in Ref.~\cite{Sumino:2010bv}.
Moreover, the differential cross section derived from the Green function method is only reliable around the $m_{t\bar{t}}=2m_t$ threshold, where the non-relativistic assumption made in Eq.~\eqref{eq:Green} is justified. We follow the upper bound $m_{t\bar{t}}\,<\,$350~GeV used in several studies based on non-relativistic QCD (NRQCD) to truncate our calculation, as suggested in Refs.~\cite{Fuks:2024yjj,Garzelli:2024uhe}.  To extend the applicable window to larger $m_{t\bar{t}}$, we refer to Ref.~\cite{Sumino:2010bv}. 
In general, one can also generalize the Green function method to the $P$-wave case; however, as we will demonstrate in Section~\ref{subsec:chi_t}, the production cross sections of $P$-wave toponia at future colliders are too small to be sensitive to experimental searches, and thus we only present the corresponding comparison in Appendix~\ref{subapp:GF:diff}.

While a detailed review and rigorous application of the Green function method is beyond the scope of this study, we note here some caveats of the method that are still under active research (see for example Refs.~\cite{Kuhn:1992qw,Kiyo:2008bv,Beneke:2011mq,Beneke:2015kwa,Ju:2020otc,Garzelli:2024uhe}). Despite the intention of the Green function method to incorporate bound-state effects into the calculation of the $m_{t\bar{t}}$ spectrum, there exist inevitable scattering-state contributions to its final outcome that need to be properly subtracted and matched. There are two main reasons behind this~\cite{Hagiwara:2008df}: 1. While in the $\Gamma_t\to0$ limit, one can safely rely on the Green function method to resum the total cross section close to the threshold expanded in $(\alpha_s/v)^n$~\cite{Hoang:2000yr}, $v$ being the top quark velocity in the $t\bar{t}$ rest frame, in reality because of the large $\Gamma_t$, a broad ``resonance'' across the entire threshold regime will receive such corrections. 2. Unique to hadron colliders, additional radiation of gluons becomes significant near the threshold and can be resummed in terms of the expansion in $(\alpha_s^2\ln^2v)^n$, which should also be taken into account when performing higher-order corrections to both the Coulomb-gluon exchange effects and the fixed-order calculations~\cite{Hagiwara:2008df}. Consequently, it is important to match the results of the Green function method and fixed-order calculations to properly distinguish between the bound-state and the scattering-state contributions to the total cross section near the threshold. Before then, one cannot claim with certainty whether the $t\bar{t}$ threshold enhancements measured by CMS and ALTAS actually imply the existence of $\eta_t$.

\section{Sensitivity of toponium states at the LHC and future colliders}\label{sec:sensitivity}

In this section, we discuss the discovery sensitivity of $\eta_t$, $\psi_t$, $\chi_{t0}$, and $\chi_{t1}$ at the LHC and future colliders, including the HL-LHC and FCC-ee. 
For the associated annihilation decays and production cross sections, we calculate them using the projection method reviewed in Sections~\ref{subsec:decay} and \ref{subsec:prod} and ignore the $K$-factors unless otherwise specified. A brief summary of the proposed specs of the  mentioned future colliders is given in Table~\ref{tab:colliders}.

\begin{table}[ht!]
    \centering
    \renewcommand{\arraystretch}{1.25}
    \begin{tabular}{c||c|c}
        \hline\hline
        future collider & $\sqrt{s}$ & expected integrated luminosity ($\mathcal{L}$) \\
        \hline
        HL-LHC & $14$~TeV & $3~{\rm ab}^{-1}$ \\
        \hline
        \multirow{2}{*}{FCC-ee} & $360$~GeV & $2.70~{\rm ab}^{-1}$ \\
        & $340$-$350$~GeV & $0.42~{\rm ab}^{-1}$ \\
        \hline\hline
    \end{tabular}
    \renewcommand{\arraystretch}{1}
    \caption{A brief summary of the proposed specs of the HL-LHC~\cite{CMS:2025hfp} and FCC-ee~\cite{FCC:2025lpp,FCC:2025uan}.
    }
    \label{tab:colliders}
\end{table}

\subsection{$\eta_t$}\label{subsec:eta_t}

We begin our discussion of $\eta_t$ with the recent measurement performed by CMS in Ref.~\cite{CMS:2025kzt}, which assumes $m_t=m_t^{\rm MC}=172.5~{\rm GeV}$ and $M_{\eta_t}=343~{\rm GeV}$ (and thus implies $E_{\rm bind}=-2$~GeV) and reports $\sigma(pp\to\eta_t)=8.8\substack{+1.2\\-1.4}~{\rm pb}$. ATLAS also performed the same study and reported $\sigma(pp\to\eta_t)=9.3\substack{+1.4\\-1.3}~{\rm pb}$~\cite{ATLAS:2026dbe}. In both studies, the quasi-bound-state mass $M_{\eta_t}$ is treated as a fixed value rather than being fitted from the data.
Both collaborations have in fact observed an excess in $t\bar{t}$ production near the threshold region. As shown in Figure~\ref{fig:xs_pp_etat_GreenFunc} and calculated more precisely in Ref.~\cite{Beneke:2011mq}, the signal receives contributions from both quasi-bound states and the continuum. However, the current experimental analyses use bins of the $t\bar{t}$ invariant mass that are too large to resolve these two components separately. On the other hand, the observed excess provides a strong hint for the existence of a quasi-bound state near the threshold, given the comparable predicted magnitude of the toponium contribution within the NRQCD framework.

\begin{figure}[th!]
    \centering
    \includegraphics[width=0.6\linewidth]{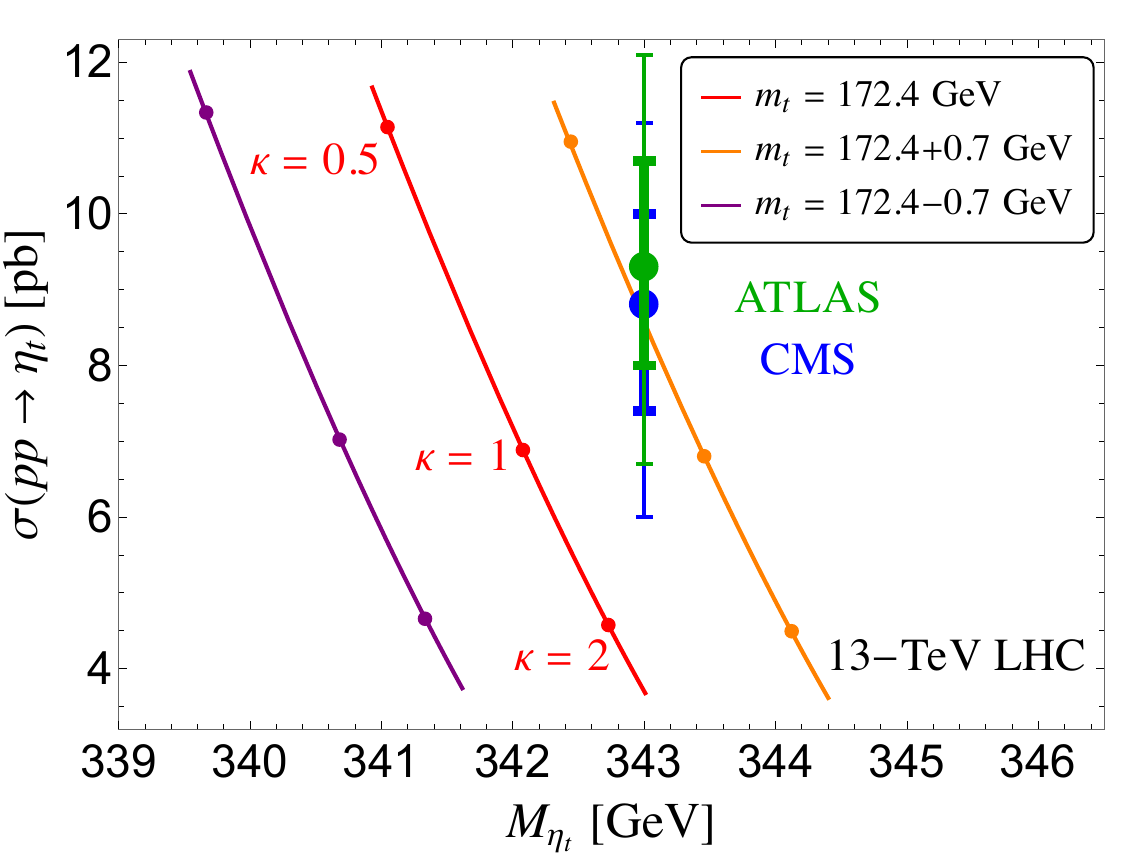}
    \caption{The theoretical contours on the $M_{\eta_t}$--$\sigma(pp\to\eta_t)$ (13-TeV LHC) plane derived from varying $\kappa\equiv \mu_R\times r$ with $m_t=172.4~{\rm GeV}$ (red), $m_t=172.4+0.7~{\rm GeV}$ (orange), and $m_t=172.4-0.7~{\rm GeV}$ (purple), respectively, according to the pole mass $m_t^{\rm pole}=172.4\pm0.7$~GeV reported in Ref.~\cite{ParticleDataGroup:2024cfk}. We also present the experimental results reported in Ref.~\cite{CMS:2025kzt} (blue) and Ref.~\cite{ATLAS:2026dbe} (green), respectively, at $1\sigma$ (thick lines) and 95\% CL (thin lines) assuming $m_t=m_t^{\rm MC}=172.5$~GeV and $M_{\eta_t}=343$~GeV. We neglect the $K$-factor since it is close to one as discussed in Ref.~\cite{Sumino:2010bv}. 
    }
    \label{fig:BindE_XS_pp_etat}
\end{figure}

To see what the experimental results might imply, we show the contours derived from varying $\kappa\equiv \mu_R\times r$ on the $M_{\eta_t}$--$\sigma(pp\to\eta_t)$ plane with $m_t$ chosen to be $172.4$~GeV, $172.4+0.7$~GeV, and $172.4-0.7$~GeV according to the the measured top quark pole mass reported in Ref.~\cite{ParticleDataGroup:2024cfk}~\footnote{
For a detailed discussion of different definitions of the top quark mass and how they can be measured from experiments, see Ref.~\cite{CMS:2024irj}.} in Figure~\ref{fig:BindE_XS_pp_etat}, in which we also show the experimental results at $1\sigma$ and 95\% confidence level (CL). Again, we ignore the $K$-factor since it is close to one as discussed in Ref.~\cite{Sumino:2010bv}.
One can see that the theoretical contours cover a large area of the $M_{\eta_t}$--$\sigma$ plane under the variation of $m_t$ and $\kappa$, signaling that the experimental result should be further improved to properly reflect these uncertainties in the predicted value of $M_{\eta_t}$. 
Moreover, while the theoretical predictions here are obtained with the MSTW 2008 NNLO PDF~\cite{Martin:2009iq}, it is also important to consider the uncertainties arising from different PDF choices, although we have checked that most will agree within systematic errors.
Therefore, it remains an open issue for experimental searches to properly account for the underlying theoretical uncertainties in the $\eta_t$ measurement to test the SM QCD, which could then later be used to more rigorously probe certain BSM physics, as we will discuss with an explicit example in Section~\ref{sec:SSM}. 

\begin{table}[hb!]
    \centering
    \renewcommand{\arraystretch}{1.5}
    \begin{tabular}{c|c|c|c|c}\hline\hline
        process & $\eta_t \to ZH$ & $Zb\bar{b}$ & $t\bar{t}$ & $ZH$ \\ \hline \hline
        cross section$\times$BR [fb] & 0.32 & $7.52\times10^{4}$ & $2.76\times10^{4}$ & 22.47 \\ \hline
        $Z, H$ reconstruction  & 7.87\% & 0.15\%  & 1.81\% & 4.91\% \\ \hline
        $E^{\text{miss}}_T/\sqrt{H_T}<1.15\sqrt{{\rm GeV}}$  & 4.28\% & 0.07\% & 0.10\% & 2.99\%\\ \hline
        $310 <m_{ZH}/\mbox{GeV} < 370$  & 3.81\% & 0.01\% & 0.03\% & 0.53\% \\ \hline\hline
        fiducial cross section [fb] & 0.012 & 6.775 & 8.280 & 0.118\\ \hline\hline
    \end{tabular}
    \renewcommand{\arraystretch}{1}
    \caption{The signal and backgrounds of the search for $pp \rightarrow \eta_t \rightarrow (Z \rightarrow \ell^+ \ell^-)(H \rightarrow b\bar{b})$. Note that the backgrounds coming from the dileptonic decay of the continuum $t\bar{t}$ and the dominant $Zb\bar{b}$ from $Z+$heavy flavors are included following Ref.~\cite{ATLAS:2022enb}.}
    \label{tab:eff_tab_etaZH}
\end{table}

Given the strong hint of the existence of $\eta_t$ in the gluon fusion channel via the individual top decays, it is valid to ask what other avenues we can use to study its properties. One possibility is to probe its next-to-leading annihilation decay to $ZH$, which has a branching ratio of roughly $1.02\times10^{-3}$. Along with the measured production cross section $\sigma(pp\to\eta_t)=8.8$~pb, one obtains $\sigma(pp\to\eta_t\to ZH)=8.98$~fb at the 13-TeV LHC. In the following, we perform the Monte Carlo studies, including those for the corresponding backgrounds, using \texttt{MadGraph5\_aMC@NLO}~\cite{Alwall:2014hca} at the parton level, \texttt{Pythia 8}~\cite{Sjostrand:2014zea} for parton showering and hadronization, and \texttt{Delphes 3}~\cite{deFavereau:2013fsa} for detector effects.
We use the model provided in Ref.~\cite{Fu:2025zxb} to simulate the processes and reconstruct the $Z$ boson from the dileptonic events $\ell^+\ell^-$ and the $H$ boson from $b\bar{b}$. In this channel, the main backgrounds come from the continuum $Zb\bar{b}$ and the dileptonic decay of $pp\to t\bar{t}$~\cite{ATLAS:2022enb}. In addition, the production of continuum $pp\to ZH$ is also considered. The cuts applied in the analysis are as follows, with their efficiencies reported in Table~\ref{tab:eff_tab_etaZH}: We select the events that satisfy the basic cuts on the transverse momenta and pseudo-rapidities, $p_{T}^{\ell,j}>25~\mathrm{GeV}$ and $|\eta_{j,\ell}|<2.5$. For $ZH$ reconstruction, we require at least two $b$-jets and two leptons in the final state. The $Z$ boson is reconstructed by requiring $77~{\rm GeV}<m_{\ell\ell}<101.2~{\rm GeV}$, and the $H$ boson $100~{\rm GeV}<m_{b\bar{b}}<145~{\rm GeV}$, following Ref.~\cite{ATLAS:2022enb}. To suppress the $t\bar{t}$ background, we impose a missing transverse energy cut $E^{\text{miss}}_T/\sqrt{H_T}<1.15\sqrt{{\rm GeV}}$, where $H_T$ is the scalar sum of the transverse momenta of the leptons and $b$-jets. We then select the invariant mass window $310~{\rm GeV}<m_{ZH}<370~{\rm GeV}$ to enhance the signal-to-background contrast. Consequently, assuming an integrated luminosity of $3~{\rm ab}^{-1}$ at the HL-LHC, we find the significance to be only $0.17\sigma$, which is mainly suppressed by the large backgrounds coming from $Zb\bar{b}$ and $t\bar{t}$. Assuming the same efficiencies, we estimate that the $2\sigma$ exclusion upper bound on the production cross section for a pseudoscalar boson decaying into $ZH$ is around $500~{\rm fb}$, consistent with the $\sim300~{\rm fb}$ limit shown in Ref.~\cite{ATLAS:2022enb} which additionally includes the $Z\to\nu\nu$ channel. Moreover, the $\eta_t\to W^+W^-$ decay channel is also highly suppressed according to the $W^+W^-$ resonance searches to date. For example, Ref.~\cite{ATLAS:2023alo} reports a sensitivity of around $500~{\rm fb}$ for the resonance production of $W^+W^-$ at $350~{\rm GeV}$, while the predicted $\sigma(pp\to\eta_t\to W^+W^-)$ is only about $2~{\rm fb}$. Therefore, we conclude that neither $\eta_t\to ZH$ nor $\eta_t\to WW$ can be probed at the current LHC or the future HL-LHC.

\begin{figure}[ht!]
    \centering
    \subfigure[$gg\to\eta_t b\bar{b}$]{\includegraphics[width=0.48\linewidth, trim= 2cm 19cm 2cm 2.5cm, clip]{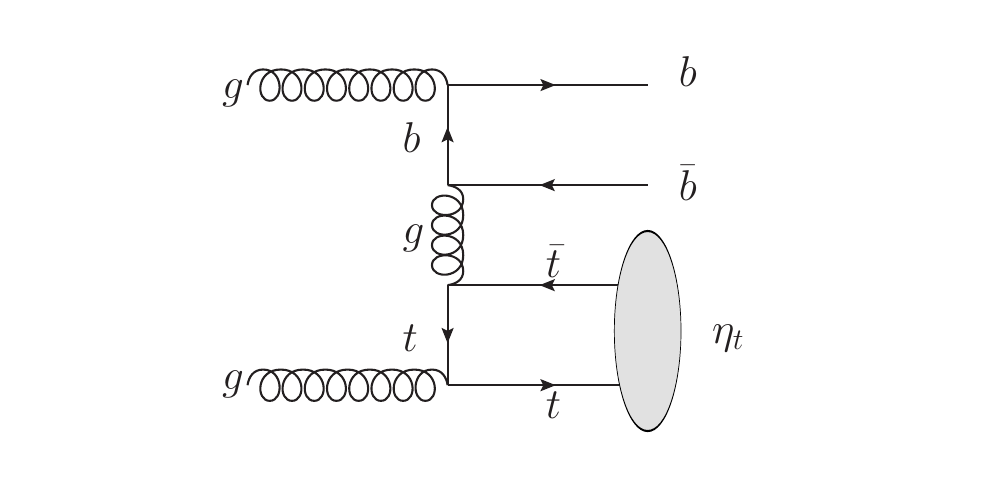}\label{fig:FeynDiag_gg_etatbb}}
    \quad
    \subfigure[$e^+ e^-\to\eta_t\gamma/\psi_t\gamma$]{\includegraphics[width=0.48\linewidth, trim= 2cm 19cm 2cm 2.5cm, clip]{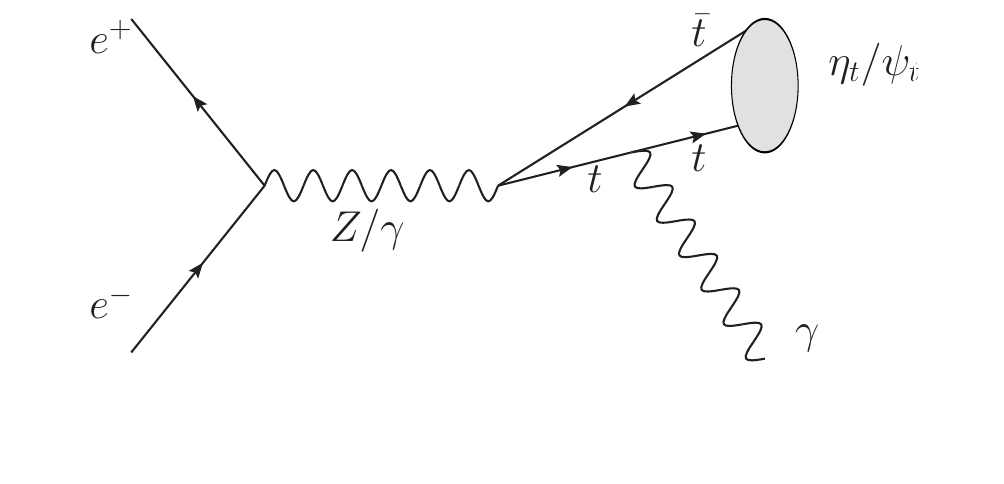}\label{fig:FeynDiag_ee_etatgamma}}
    \quad
    \subfigure[$gg\to\psi_t b\bar{b}$]{\includegraphics[width=0.48\linewidth, trim= 2cm 19cm 2cm 2.5cm, clip]{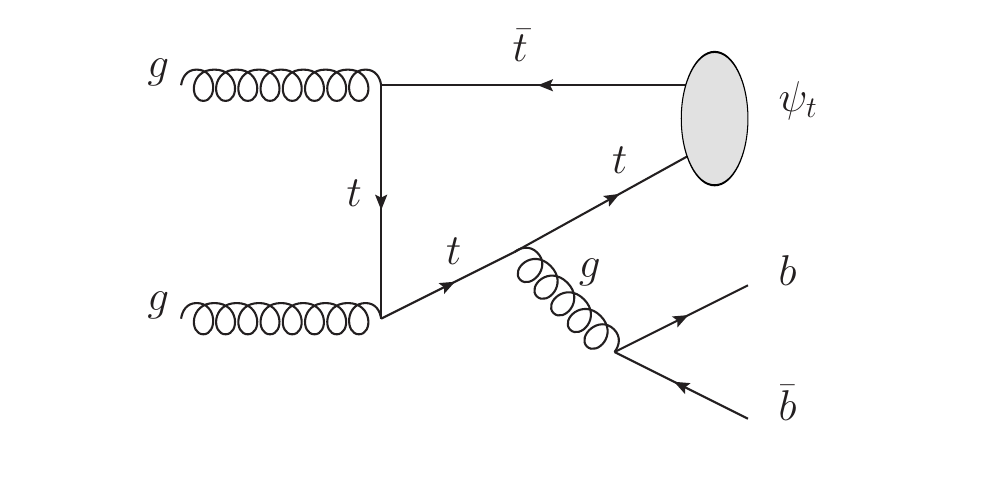}\label{fig:FeynDiag_gg_psitbb}}
    \quad
    \subfigure[$q\bar{q}^\prime\to\psi_t W$]{\includegraphics[width=0.48\linewidth, trim= 2cm 18cm 2cm 2.5cm, clip]{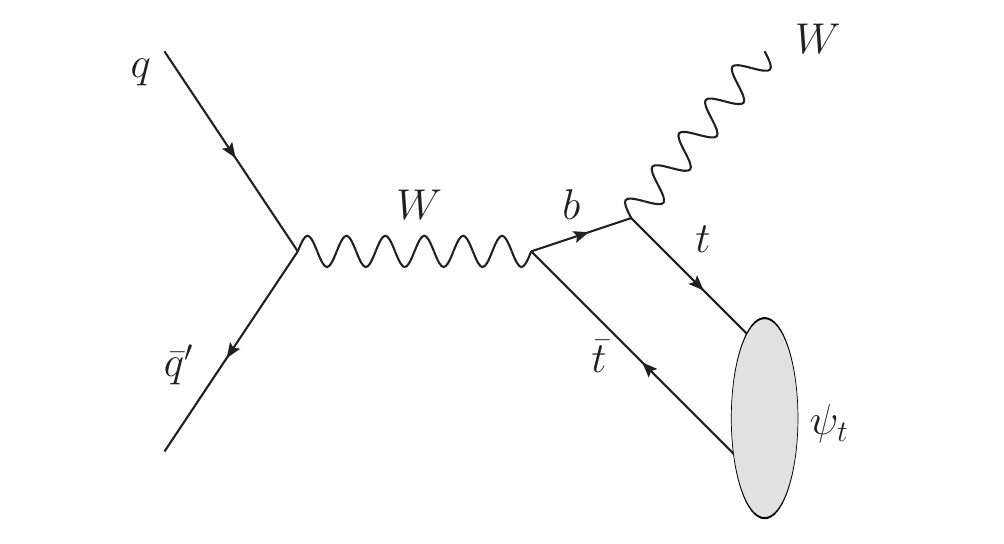}\label{fig:FeynDiag_ud_psitW}}
    \caption{Example Feynman diagrams of the specified processes considered in this study.}
    \label{fig:feynman_diag}
\end{figure}

Another promising channel is $gg\to \eta_t b\bar{b}$, an example Feynman diagram of which is shown in Figure~\ref{fig:FeynDiag_gg_etatbb}. Taking into account only the QCD-induced diagrams and using the projection method, one obtains $\sigma(pp\to \eta_t b\bar{b})=0.37~{\rm pb}$ at the 13-TeV LHC. Compared to the continuum background with $\sigma(pp\to t\bar{t}b\bar{b})=16.3~{\rm pb}$ calculated using Monte Carlo simulations at tree level, the ratio between them is of the same order of magnitude as that in the direct $\eta_t$ search. Therefore, this process is likely to be probed at both the current LHC and the future HL-LHC, which we explore in the dileptonic decay channel of the top quark pairs in what follows.

The $t\bar{t}b\bar{b}$ channel has been studied at the LHC (see for example Ref.~\cite{ATLAS:2024aht}). Here we use a simplified pseudoscalar model for $\eta_t$ to simulate the signal kinematics but stick to the cross section calculated using the projection method reported above in order to avoid potential issues related to the effective vertices, as we will discuss in Section~\ref{sec:conclusions}. We apply the basic cuts $p_{T}^{\ell,j}>25~{\rm GeV}$ and $|\eta_{j,\ell}|<2.5$ and require at least two leptons and four jets in the final state, with at least three of the jets tagged as $b$-jets. The neutrino momenta are reconstructed using four-momentum conservation together with the on-shell conditions for $W$ and $t$, after which we select from the possible solutions the one that gives the smallest $m_{t\bar{t}}$. Then, we impose the mass window cut $330~{\rm GeV}<m_{t\bar{t}}<370~{\rm GeV}$ around the $\eta_t$ mass. Finally, we also study the angular observables $c_{\rm hel}$ and $c_{\rm han}$ defined in the CMS and ATLAS studies for $t\bar{t}$ threshold production~\cite{CMS:2025kzt,ATLAS:2026dbe} and find that the cut $c_{\rm hel}>1/3$ improves the significance by only about $4\%$, which we still include in our analysis.

\begin{figure}[th!]
    \centering
    \includegraphics[width=0.5\linewidth]{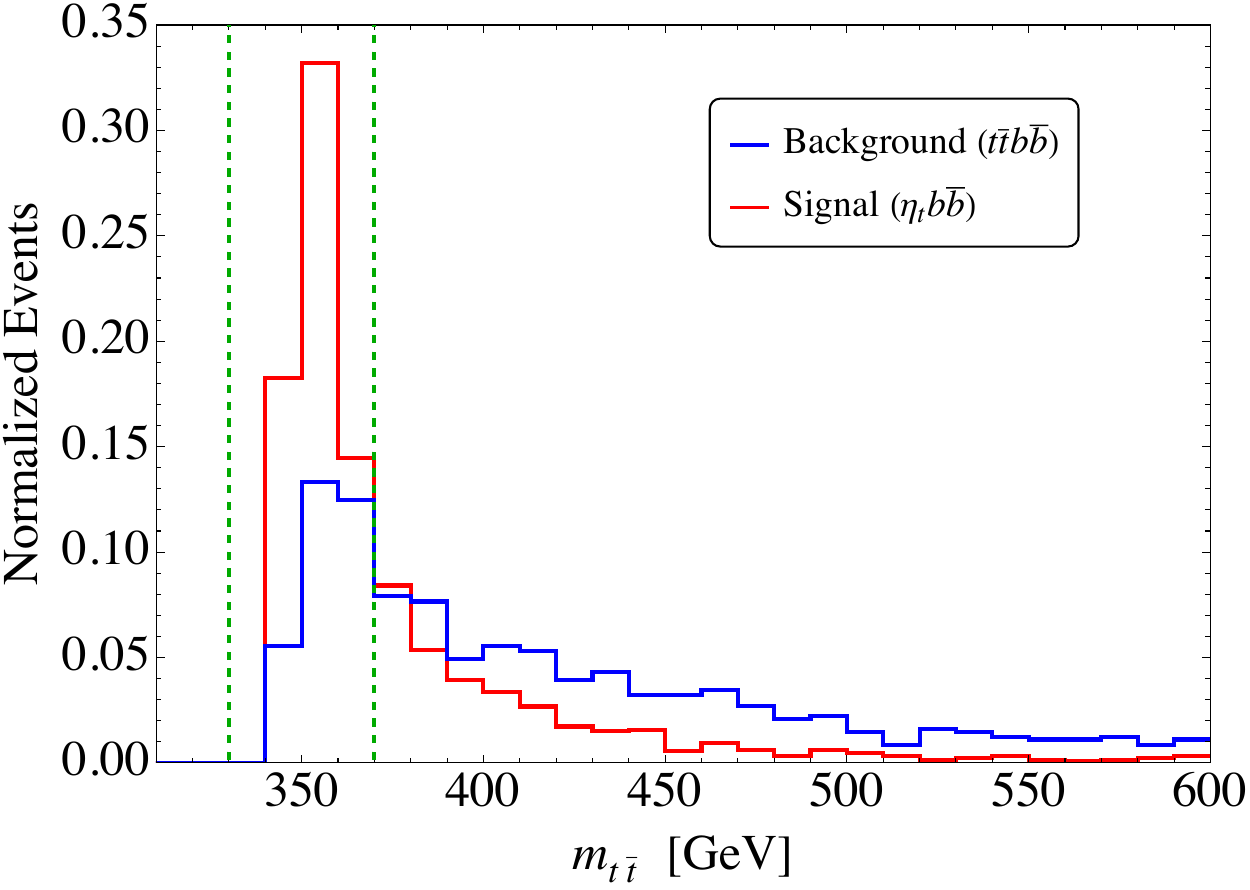}
    \caption{The normalized $m_{t\bar{t}}$ distributions obtained from the reconstructed $t\bar{t}$ events for the $\eta_t b\bar{b}$ signal (red) and the $t \bar{t}b\bar{b}$ background (blue). The green dashed lines indicate the selection window $330~{\rm GeV}<m_{t\bar{t}}<370~{\rm GeV}$.}
    \label{fig:recon_ttbar_etatbb}
\end{figure}

The signal and background cross sections and efficiencies are summarized in Table~\ref{tab:eff_tab_etabb}, and the reconstructed $m_{t\bar{t}}$ distributions are shown in Figure~\ref{fig:recon_ttbar_etatbb}.
Together with the dileptonic branching ratios and including the $K$-factor of $1.77$~\cite{Pan:2019wwv}, the significance is estimated to be about $6.1\sigma$ at the HL-LHC, making it a promising channel that can serve as complementary evidence that the pseudoscalar resonance measured at the LHC indeed originates from the quasi-bound state effects of the $t\bar{t}$ system, rather than from a new-physics pseudoscalar $A$, such as in the two-Higgs-doublet model. A new-physics state $A$ can have an $Ab\bar{b}$ coupling that is independent of QCD dynamics, and the process $pp\to Ab\bar{b}$ can be produced through diagrams involving this $b$-quark coupling. In contrast, $pp\to \eta_t b\bar{b}$ proceeds through gluon couplings and is predominantly determined by QCD interactions, up to minor electroweak contributions that we find negligible from our calculations. Therefore, if the HL-LHC measures a cross section consistent with the predicted value, it would provide strong evidence that the pseudoscalar resonance associated with the $t\bar{t}$ system indeed exists.

\begin{table}[hb!]
    \centering
    \renewcommand{\arraystretch}{1.25}
    \begin{tabular}{c|c|c}\hline\hline
        process & $\eta_t b\bar{b}$ & $(t\bar{t}b\bar{b})_{\rm SM}$ \\ \hline\hline
        cross section$\times$ BR [fb] & 16.8 & 1313\\ \hline
        basic selection  & 17.75\% & 15.00\%\\ \hline
        $t,\bar{t}$ reconstruction  & 2.57\% & 1.81\%\\ \hline
        $330 < m_{t\bar{t}}/\mbox{GeV} < 370$ & 1.68\% & 0.51\%\\ \hline
        $c_{\rm hel} > 1/3$  & 1.00\% & 0.16\%\\ \hline\hline
        fiducial cross section [fb] & 0.17 & 2.11 \\ \hline\hline
    \end{tabular}
    \renewcommand{\arraystretch}{1}
    \caption{The cross sections and efficiencies obtained from the study of $\eta_t b \bar{b}$ production, including those of the signal and the continuum background $t\bar{t}b \bar{b}$. The branching ratio (BR) used here for both the signal and background corresponds to that of the dileptonic decay of $t\bar{t}$.}
    \label{tab:eff_tab_etabb}
\end{table}

On the other hand, $\eta_t$ is less likely to be directly produced at lepton colliders due to its $J^{PC}$ property, which prohibits its singular production through the process $e^+e^-\to\eta_t$ given the tiny electron mass. Some other possible production mechanisms include $e^+ e^-\to \eta_t \gamma$, $e^+ e^-\to \eta_t Z$, and $e^+ e^-\to \eta_t H$ [see Figure~\ref{fig:FeynDiag_ee_etatgamma} for an example Feynman diagram of $e^+e^-\to\eta_t \gamma$], which we explore using the projection method in what follows. Since the currently proposed $\sqrt{s}$ of the future lepton colliders listed in Table~\ref{tab:colliders} are all only slightly above the $t\bar{t}$ threshold, we only consider the $e^+ e^-\to \eta_t \gamma$ process. Taking $\sqrt{s}=360~\text{GeV}$, which is proposed for the FCC-ee, one obtains $\sigma(e^+e^-\to\eta_t\gamma)=8.87\times10^{-3}~{\rm fb}$ after applying the photon cuts $p_{T}^{\gamma}>10$~GeV and $\vert\eta_{\gamma}\vert<3.0$. Compared to the continuous background with the same cuts applied, which has $\sigma(e^+e^-\to t\bar{t}\gamma)=8.57$~fb according to Monte Carlo simulations, it is clear that $\eta_t$ is unlikely to be probed via this process and thus at lepton colliders in general. 

\subsection{$\psi_t$}\label{subsec:psi_t}

For the $\psi_t$ state, since it cannot be produced singularly through gluon fusion as prohibited by the Landau-Yang theorem~\cite{Landau:1948kw,Yang:1950rg}, we consider other (associated) production mechanisms: $gg\to\psi_t g$, $gg\to \psi_t b\bar{b}$, and $q\bar{q}^\prime \to \psi_t W$ at the LHC and $e^+e^-\to\psi_t$ at lepton colliders. For the first process, the parton-level associated production cross section is given by
\begin{equation}
\begin{aligned}
    \hat{\sigma}(gg\to\psi_t g) &= \frac{10\pi\alpha_s^3}{9M\hat{s}^2(\hat{s}-M^2)^2(\hat{s}+M^2)^3}\vert R_S(0)\vert^2\bigg[
        (\hat{s}^2-M^4)(2\hat{s}^3+M^2\hat{s}^2+4M^4\hat{s}+M^6) \\
        &\quad - 2M^4\hat{s}(M^4+2M^2\hat{s}+5\hat{s}^2)\log\left(\frac{\hat{s}}{M^2}\right)
    \bigg] ~,
\end{aligned}
\end{equation}
which gives the proton-level cross section $\sigma(pp\to\psi_t g)=73.3~{\rm fb}$ at the 13-TeV LHC. Here, $M=M_{\psi_t}$. This result is overwhelmed by the continuous production cross section $\sigma(pp\to t\bar{t}+j)=\mathcal{O}(10^3)~{\rm pb}$. For the second process, the cross section given by the projection method using the QCD-induced diagrams [see for example Figure~\ref{fig:FeynDiag_gg_psitbb}] is $\sigma(pp\to \psi_t b \bar{b})\sim\mathcal{O}(2)~{\rm fb}$ at the 13-TeV LHC, while the electroweak contribution is $\sim\mathcal{O}(0.3)~{\rm fb}$, the combined result of which is significantly prevailed by both the continuous background with $\sigma(pp\to t\bar{t}b\bar{b})=16.3$~pb and the $\eta_t$-induced background with $\sigma(pp\to\eta_t b\bar{b})=0.37$~pb. Finally, for the third process [see for example Figure~\ref{fig:FeynDiag_ud_psitW}], the associated cross section is $\sigma(pp\to \psi_t W)=0.09~{\rm fb}$ at the 13-TeV LHC, which is again overwhelmed by the continuous background with $\sigma(pp\to t\bar{t}W)=369$~fb as reported by Monte Carlo studies. Therefore, we conclude that it is unlikely to probe $\psi_t$ at both the current LHC and the future HL-LHC.

\begin{figure}[th!]
    \centering
    \includegraphics[width=0.6\linewidth]{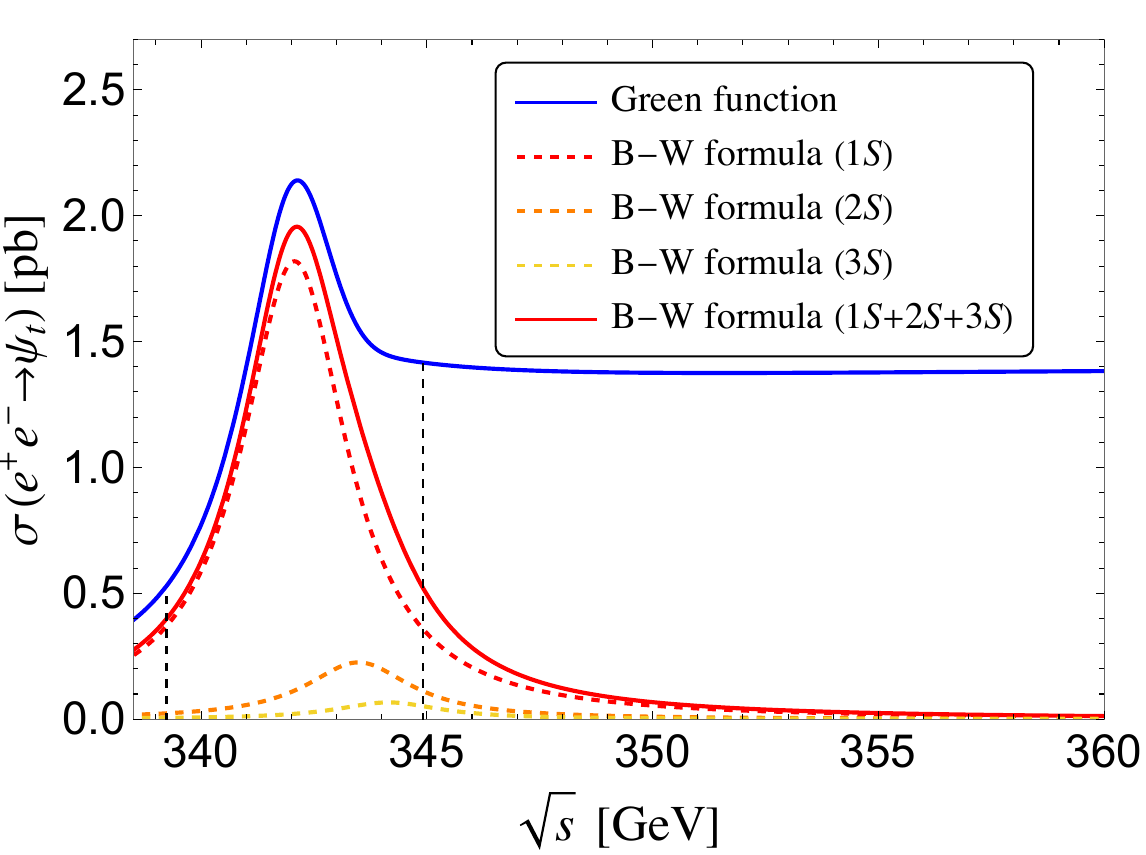}
    \caption{The cross section $\sigma(e^+e^-\to t\bar{t})$ as a function of $\sqrt{s}$ calculated using both the Green function method (blue) and the B-W formula (red). The two vertical dashed lines denote $m_{t\bar{t}}=M_{\psi_t}\pm\Gamma$, respectively, between which we identify the ``resonance'' region. We also show the individual $nS$ resonances with $n=1$ (red), $n=2$ (orange), and $n=3$ (yellow) in dashed curves, as in Figure~\ref{fig:xs_pp_etat_GreenFunc}.
    }
    \label{fig:xs_ee_psit+hardgamma}
\end{figure}

On the other hand, the production of $\psi_t$ at lepton colliders has been widely studied in the literature (see for example Refs.~\cite{Sumino:1992ai,Fujii:1993mk,Nagano:1999nw,Beneke:1999qg}). In particular, FCC-ee~\cite{FCC:2025uan,FCC:2025lpp} proposes to run at the $t\bar{t}$ threshold around $\sqrt{s}\sim360$~GeV, which would most likely further cover the mass pole of $\psi_t$ (together with other excited states) around $2m_t$, and thus we discuss the prospect of $\psi_t$ measurements at future lepton colliders. First of all, we calculate the production cross section $\sigma(e^+e^-\to\psi_t)$ around $\sqrt{s}=M_{\psi_t}$ using both the Green function method and the B-W formula in Figure~\ref{fig:xs_ee_psit+hardgamma}, based on which we further estimate the significance of $\psi_t$ discovery through its intrinsic decay to $W^+W^-b\bar{b}$ using the B-W formula. The continuous background $e^+ e^- \to W^+W^- b \bar{b}$ has a cross section of $30.5~{\rm fb}$ at $\sqrt{s}=M_{\psi_t}$ and increases rapidly after $\sqrt{s}$ passes through $2m_t$. Assuming the reconstruction efficiency of the $t\bar{t}$ final state to be above 10\%, the integrated luminosity to be $1~{\rm ab}^{-1}$, and using both the dileptonic and semi-leptonic final states from top decays, one could easily discover $\psi_t$ at $\sqrt{s}=M_{\psi_t}$.

Moreover, in the recent Ref.~\cite{Fu:2024bki}, it is claimed that the two-body annihilation decays of $\psi_t$ could be discovered with a significance of over 5\,$\sigma$ at a lepton collider by measuring the cross section ratio $R_b\equiv \sigma(e^+e^-\to b\bar{b})/\sum_{q=u,d,s,c,b}\sigma(e^+e^-\to q\bar{q})$ at a few different $\sqrt{s}$ points. The main idea is to capture the interference effect caused by the $\psi_t$-mediated diagrams when $\sqrt{s}$ passes from below $M_{\psi_t}$ to above $M_{\psi_t}$. As a result of this interference, $R_b(\sqrt{s})$, which would only be a flat line around $\sqrt{s}=M_{\psi_t}$ if only the $\gamma$- and $Z$-poles were present, will manifest an interference pattern that can facilitate the measurement of the $\psi_t$-pole. One major factor enabling this possibility is the relatively large annihilation decay width of $\psi_t\to b\bar{b}$ (see Table~\ref{tab:decay_width_projection_value}), owing to the $W$-mediated contributions, which are highly suppressed for down-type quarks of the other generations but not for bottom quarks. Nevertheless, we note that their follow-up paper, Ref.~\cite{Fu:2025zxb}, contains a sign error in the coupling $g^V_{\psi_t}$ in the interaction $\bar{b}\gamma_\mu(g^V_{\psi_t}-g^A_{\psi_t}\gamma^5)b\,\psi_t^\mu$ (their earlier paper Ref.~\cite{Fu:2024bki} and the public model file are both correct) and also uses a larger value of $\alpha_s$ than in our study.

To assess the interference effects in $\psi_t\to b\bar{b},~W^+W^-,~\gamma Z$, we perform a simple analysis in which the significance is estimated via $S/\sqrt{B}$. Here $S$ represents the deviation of the number of events from the SM prediction without the $\psi_t$ quasi-bound state, which arises mainly from interference effects. We focus on a single energy point that yields the highest significance and assume that the full luminosity of 420~${\rm fb}^{-1}$ is accumulated at that point to maximize the sensitivity.

We use the ``bound state effective field theory'' (BSEFT) developed in Ref.~\cite{Fu:2025zxb} to compute the cross sections along with Monte Carlo simulations, and the validity of this approach is discussed in Section~\ref{sec:conclusions}. The interference patterns of the cross sections for the three processes are shown in Figure~\ref{fig:xs_psit_interference}, and the formulas for $\sigma(e^+e^-\to b\bar{b},\gamma Z)$ are listed in Appendix~\ref{subapp:toponium_decay_width}. We present the relevant effective couplings as well as the projected sensitivities for each process in the following. 

\begin{figure}[th!]
    \centering
    \includegraphics[width=0.32\linewidth]{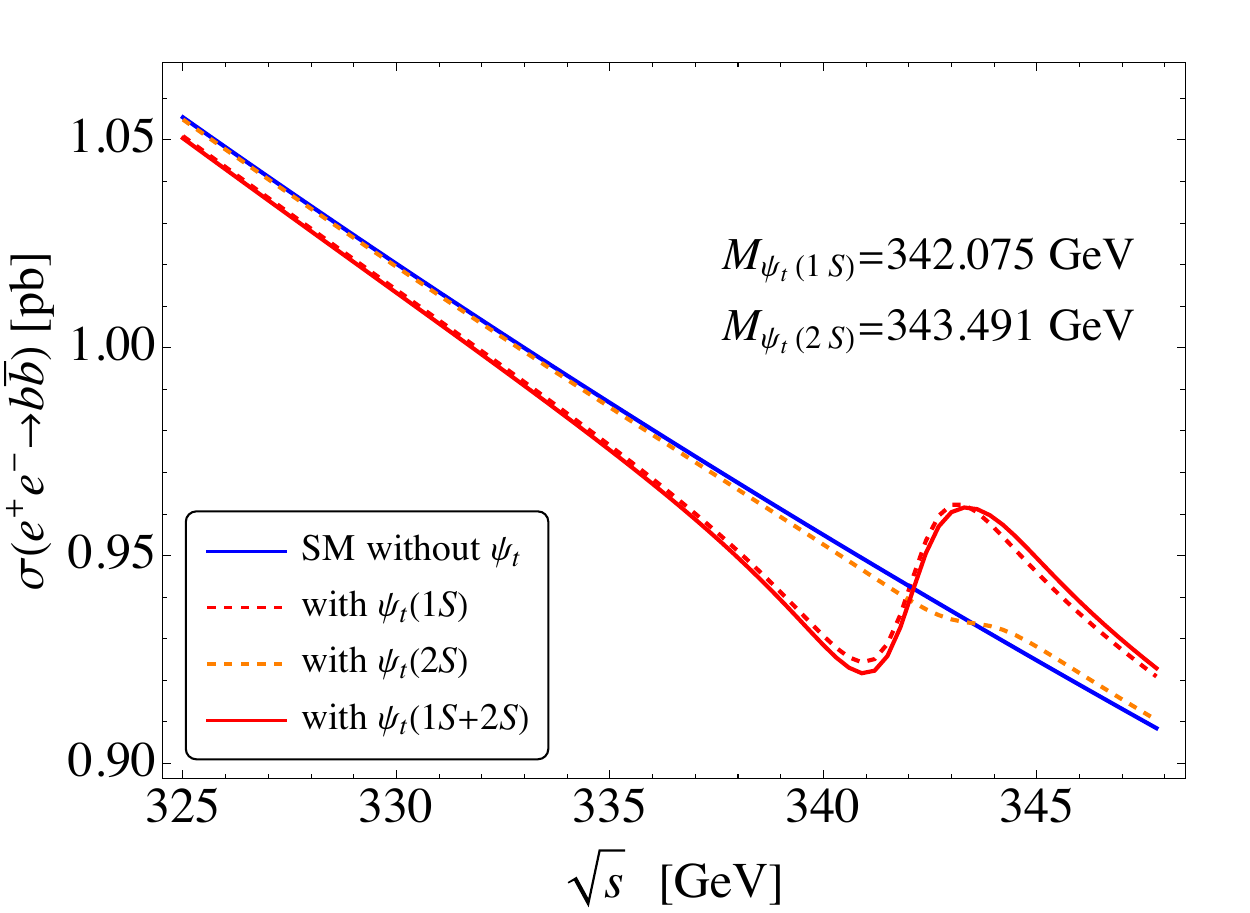}
    \includegraphics[width=0.32\linewidth]{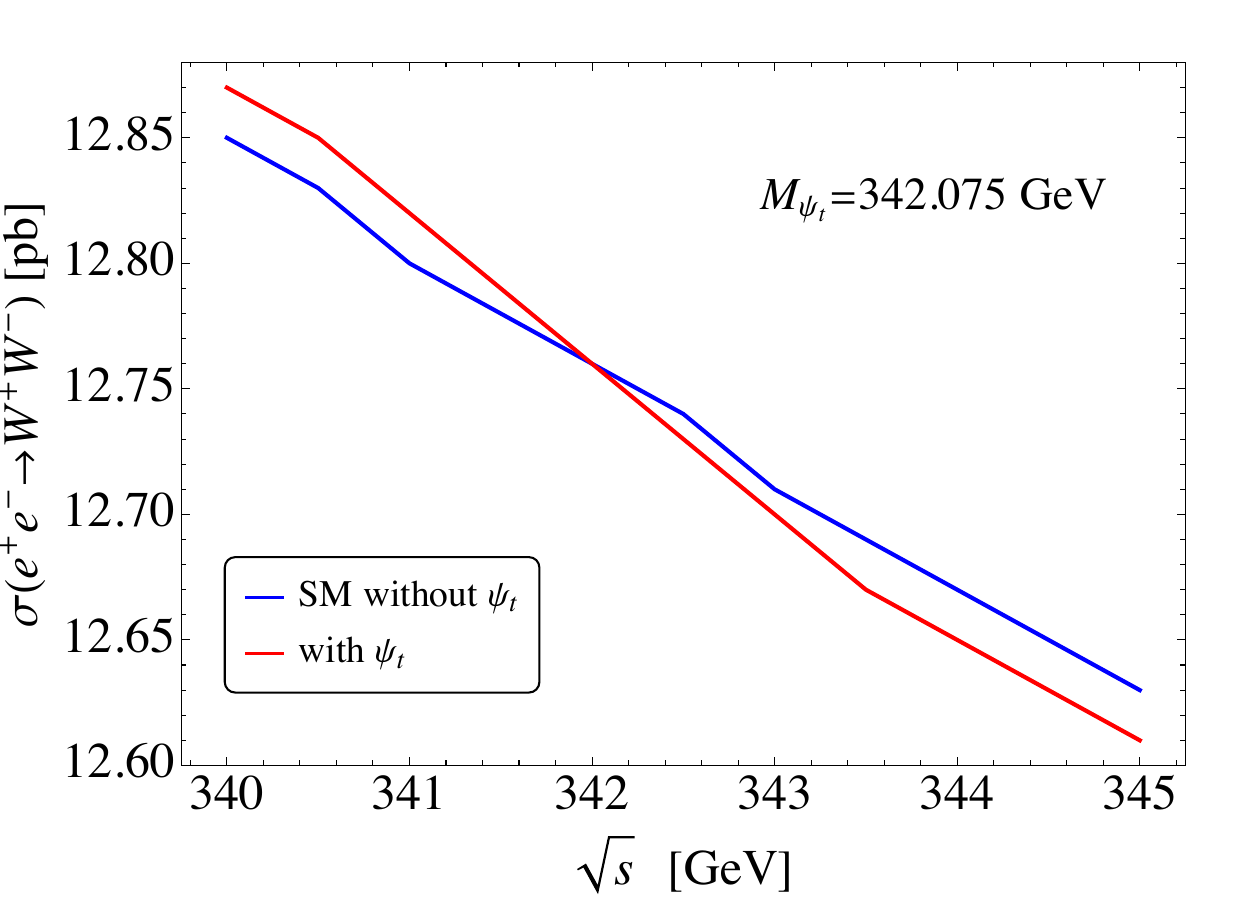}
    \includegraphics[width=0.32\linewidth]{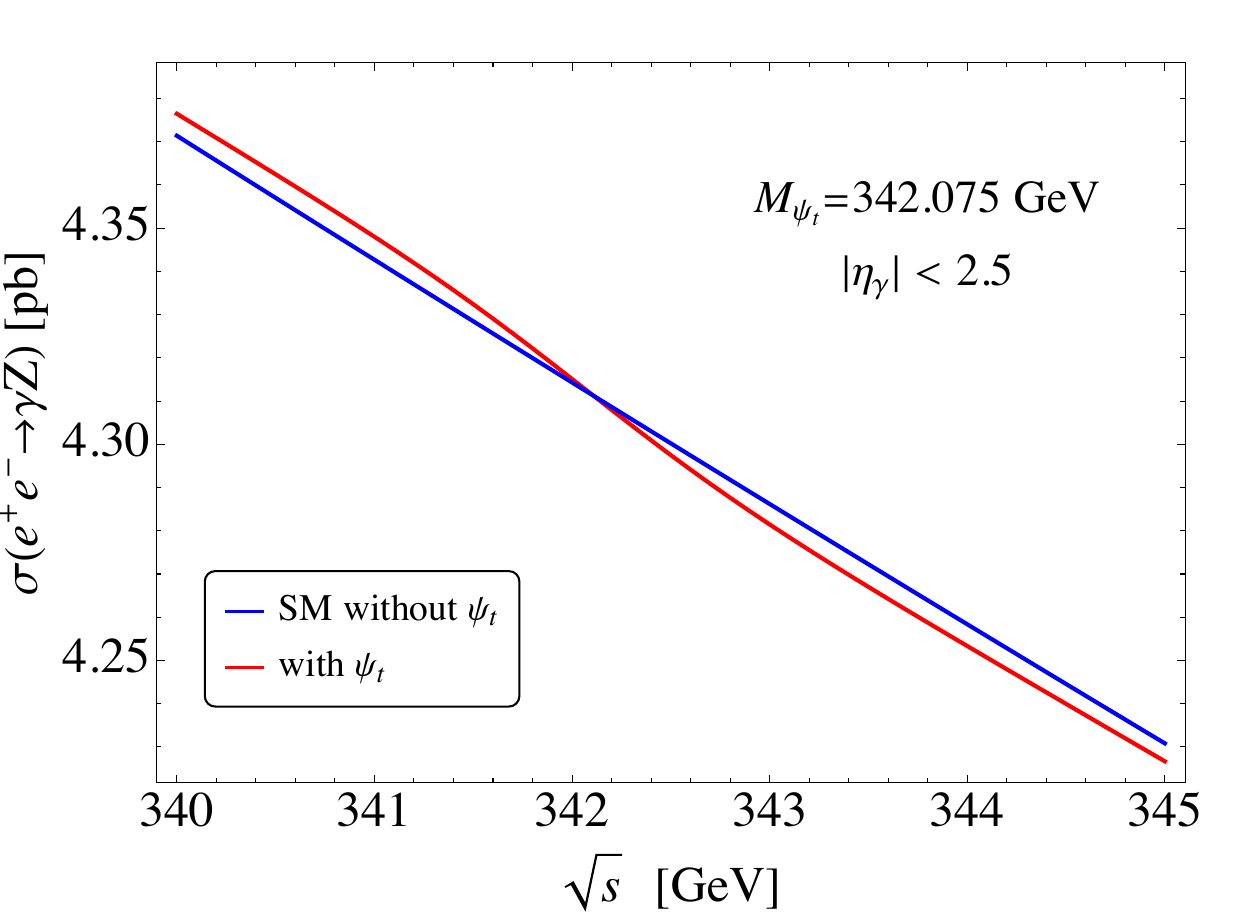}
    \caption{The cross sections of $e^+e^-\rightarrow b\bar{b}$ (left), $W^+W^-$ (middle), and $\gamma Z$ (right) predicted by the SM without including the $\psi_t$ contribution (blue) and including $\psi_t$ (red) around the $\psi_t$ mass. The $b\bar{b}$ channel shows the largest relative difference between the cases with and without $\psi_t$, so we also include the effect of $\psi_t(2S)$ for comparison. }
    \label{fig:xs_psit_interference}
\end{figure}

For the $b\bar{b}$ channel, the bound state effective couplings in $\mathcal{L}_{\psi_tbb}=\bar{b}\gamma_\mu(g^V_{\psi_t b}-g^A_{\psi_t b}\gamma^5)b \psi_t^\mu$ are
\begin{align}
    g^V_{\psi_t b}=&~\sqrt{\frac{(C_F\alpha_s)^3\pi}{3}}\frac{\alpha_{\rm EM}}{12}\left\{\frac{m_t^2}{s_w^2}\left[\frac{4(8c_w^2+1)s_w^2-9}{c_w^2(4m_t^2-m_Z^2)}-\frac{3}{m_t^2+m_W^2}-\frac{3}{m_W^2}\right]-8\right\} ~,\nonumber\\
    g^A_{\psi_t b}=&~-\sqrt{\frac{(C_F\alpha_s)^3\pi}{3}}\frac{\alpha_{\rm EM}\,m_t^2}{4s_w^2}\left[\frac{3-8s_w^2}{c_w^2(4m_t^2-m_Z^2)}+\frac{1}{m_t^2+m_W^2}+\frac{1}{m_W^2}\right] ~,\label{eqn:gAPsitBB_onshell}
\end{align}
where all three particles, $\psi_t$, $b$, and $\bar{b}$, are on shell and $M_{\psi_t}\approx 2m_t$. Here $\alpha_{\rm EM}$ is the electric fine structure constant, $c_w\equiv\cos\theta_w$ and $s_w\equiv\sin\theta_w$ with $\theta_w$ being the weak angle. In addition to the intrinsic $(b\bar{b})_{\rm SM}$ background, we also include the $(b\bar{b}jj)_{\rm SM}$ background. Other backgrounds, including mis-tagged events from $jj$ and $b\bar{b}b\bar{b}$, are negligible as shown in Ref.~\cite{Fu:2024bki}. Selected events are required to contain two $b$-tagged jets with $p_{T}^{b,j} > 20$~GeV and $\vert \eta_{b,j}\vert < 3$. The optimal $\sqrt{s}$ is found to be 343.5~GeV for $M_{\psi_t}=342$~GeV, where a significance of $14\,\sigma$ can be achieved with 420~${\rm fb}^{-1}$. The corresponding cross sections and selection efficiencies are shown in Table~\ref{tab:eff_tab_psitbb}. Given that the large significance arises from interference in this channel, we also evaluate the contribution from $\psi_t(2S)$, as shown in Figure~\ref{fig:xs_psit_interference}. It can be seen that the interference effect is dominated by the $\psi_t(1S)$ state.

\begin{table}[bh!]
    \centering
    \renewcommand{\arraystretch}{1.25}
    \begin{tabular}{c|c|c|c}\hline\hline
        process & $\psi_t\to b\bar{b}$ & $(b\bar{b})_{\rm SM}$  & $(b\bar{b}jj)_{\rm SM}$ \\ \hline\hline
        cross section$\times$ BR [fb] & 961.2 & 933.7 & 298.2 \\ \hline
        efficiency  & \multicolumn{2}{c|}{69.09\%} & 48.44\%\\ \hline
        \hline
        fiducial cross section [fb] & 664 & 645 & 144.4 \\ \hline\hline
    \end{tabular}
    \renewcommand{\arraystretch}{1}
    \caption{The cross sections and efficiencies obtained from the study of $\psi_t\to b\bar{b}$, including those of the signal and the backgrounds. Here, $M_{\psi_t}=342$~GeV and $\sqrt{s}=343.5$~GeV. }
    \label{tab:eff_tab_psitbb}
\end{table}

To search for the $\psi_t \rightarrow W^+W^-$ decay, the relevant interactions are given by
\begin{align}
    \mathcal{L}_{\psi_tWW}=&~\Bigl\{ig_{\psi_t WW,1}\left[(\partial_\sigma W_\mu^-)W^{+\sigma}-(\partial_\sigma W_\mu^+)W^{-\sigma}\right]+\frac{i}{2}g_{\psi_t WW,2}\left[(\partial_\mu W_\sigma^-)W^{+\sigma}-(\partial_\mu W_\sigma^+)W^{-\sigma}\right]\nonumber\\
    &~g_{\psi_t WW,3}\epsilon_{\mu\nu\rho\sigma}\left[(\partial^\nu W^{-\rho})W^{+\sigma}-(\partial^\nu W^{+\rho})W^{-\sigma}\right]\Bigr\}\psi_t^\mu\,,\\
    g_{\psi_t WW,1}=&~\sqrt{\frac{(C_F\alpha_s)^3\pi}{3}}\frac{\alpha_{\rm EM}}{2s_w^2}\frac{-3m_t^2(4m_b^2-4m_W^2+m_Z^2)+8m_Z^2s_w^2(m_b^2+m_t^2-m_W^2)}{(m_t^2+m_b^2-m_W^2)(4m_t^2-m_Z^2)}\,,\nonumber\\
    g_{\psi_t WW,2}=&~\sqrt{\frac{(C_F\alpha_s)^3\pi}{3}}\frac{\alpha_{\rm EM}}{s_w^2}\frac{-4m_Z^2s_w^2(m_t^2+m_b^2-m_W^2)+3m_t^2(m_Z^2+2m_b^2-2m_W^2-2m_t^2)}{(m_t^2+m_b^2-m_W^2)(4m_t^2-m_Z^2)}\,,\\
    g_{\psi_t WW,3}=&~-\sqrt{\frac{(C_F\alpha_s)^3\pi}{3}}\frac{3\alpha_{\rm EM}m_t^2}{2s_w^2(m_t^2+m_b^2-m_W^2)}\,.\nonumber
\end{align}
We consider the fully hadronic final state, which has a subleading background from $ZZ$. Jets are selected with $p_{T}^j > 20$~GeV and $\vert \eta_j \vert < 3$. Owing to the collimated nature of the two jets from the $W$ decay, we use the ``$\mbox{ee}\_\mbox{genkt}\_\mbox{algorithm}$" jet-finding algorithm with $R = 1.5$ in \texttt{Delphes 3} and impose the mass window cuts $\vert m_{j}-m_W \vert < 10$~GeV on the single-jet events and $\vert m_{jj}-m_W \vert < 10$~GeV on the dijet events. The efficiencies of these cuts are given in Table~\ref{tab:eff_tab_psitWW}. The estimated significance reaches $1.3\,\sigma$ at 340.5~GeV for a luminosity of 420\,fb$^{-1}$. With improved reconstruction efficiencies, one may observe some hints of this channel at the FCC-ee.

\begin{table}[bh!]
    \centering
    \renewcommand{\arraystretch}{1.25}
    \begin{tabular}{c|c|c|c}\hline\hline
        process & $\psi_t\to W^+ W^-$ & $(W^+ W^-)_{\rm SM}$ & $(ZZ)_{\rm SM}$ \\ \hline\hline
        cross section$\times$ BR [fb] & 5711.1 & 5702.2 & 2273.0\\ \hline
        basic selections  & \multicolumn{2}{c|}{87.77\%} & 84.79\%\\ \hline
        $W$ reconstruction  & \multicolumn{2}{c|}{33.76\%} & 10.26\%\\ \hline
        \hline
        fiducial cross section [fb] & 1924.8 & 1921.5 & 241.6 \\ \hline\hline
    \end{tabular}
    \renewcommand{\arraystretch}{1}
    \caption{The cross sections and efficiencies obtained from the study of $\psi_t\to W^+W^-$, including those of the signal and the backgrounds. The branching ratios (BRs) here all refer to the hadronic decay branching ratios. Here, $M_{\psi_t}=342$~GeV and $\sqrt{s}=340.5$~GeV. }
    \label{tab:eff_tab_psitWW}
\end{table}

For $\psi_t\to \gamma Z$ channel, the effective Lagrangian is $\mathcal{L}_{\psi_tZA}=g_{\psi_t ZA}\epsilon_{\mu\nu\rho\sigma}A^{\mu\nu}Z^\rho \psi_t^{\sigma}$, where
\begin{align}
    g_{\psi_t ZA}=\sqrt{\frac{(C_F\alpha_s)^3\pi}{3}}\frac{4\,\alpha_{\rm EM}\, m_t^2}{c_w s_w(4m_t^2-m_Z^2)}\,.\label{eqn:gPsitZA_onshell}
\end{align}
We examine two reconstruction strategies: the invisible decay $Z\to \nu\bar{\nu}$ and the hadronic $Z$ decay. For the neutrino channel, the continuous background $\gamma \nu\bar{\nu}$ away from the $Z$-pole is negligible. The reconstructed $Z$ four-momentum, $p_{Z,\text{recon}}$, is determined from energy conservation via $E_Z=\sqrt{s}-E_{\gamma}$. The basic cut on the photon is $p_{T}^\gamma > 10$~GeV and $\vert \eta_\gamma\vert < 2.5$. We impose two additional requirements: the reconstructed $Z$ mass must satisfy $\vert m_{Z,\text{recon}}-m_Z\vert < 8$~GeV, and the most energetic photon is required to have $p_{T}^\gamma > 50$~GeV.
For the hadronic $Z$ channel, we select jets with $p_{T}^j > 20$~GeV and $\vert \eta_j\vert < 3$. As in the $W^+W^-$ case, the collimation of the $Z$ decay products for some boosted $Z$ bosons makes jet identification challenging. We therefore apply both single-jet and dijet mass window cuts, $\vert m_{j}-m_Z \vert < 10$~GeV and $\vert m_{jj}-m_Z \vert < 10$~GeV.
The maximum significance occurs at $\sqrt{s}=344$~GeV. The corresponding cross sections and efficiencies are summarized in Table~\ref{tab:eff_tab_psitZgamma}. Assuming that the FCC-ee operates at this energy with $420~{\rm fb}^{-1}$, the expected significances are $0.54\,\sigma$ for the invisible $Z$ decay channel and $0.80\,\sigma$ for the hadronic decay channel.

\begin{table}[hb!]
    \centering
    \renewcommand{\arraystretch}{1.25}
    \begin{tabular}{c|c|c||c|c}\hline\hline
        process & $\psi_t\to\gamma Z_{\nu\bar{\nu}}$ & $(\gamma Z)_{\rm SM}$   & $\psi_t\to\gamma Z_{\rm hadron}$ & $(\gamma Z)_{\rm SM}$ \\ \hline\hline
        cross section$\times$ BR [fb] & 851.6 & 850.6 &  2977 & 2973 \\ \hline
        basic selections  & \multicolumn{2}{c||}{98.77\%} & \multicolumn{2}{c}{86.35\%}\\ \hline
        $Z$ reconstruction  & \multicolumn{2}{c||}{87.32\%}  & \multicolumn{2}{c}{55.06\%}\\ \hline
        $\eta_{\gamma}$ cut efficiency  & \multicolumn{2}{c||}{59.41\%}  & \multicolumn{2}{c}{36.89\%} \\ \hline\hline
        fiducial cross section [fb] & 505.927 & 505.333  & 1098.22 &1096.93 \\ \hline\hline
    \end{tabular}
    \renewcommand{\arraystretch}{1}
    \caption{The cross sections and efficiencies obtained from the study of $\psi_t\to \gamma Z$, including those of the signal and the backgrounds. The branching ratio (BR) corresponds to that of the neutrino decay on the left side and that of the hadronic decay on the right side, for both $\gamma Z$ production with and without $\psi_t$. The other background, $\gamma \nu \bar{\nu}$ without the $Z$-pole contribution, is found to be negligible. Here, $M_{\psi_t}=342$~GeV and $\sqrt{s}=344$~GeV.}
    \label{tab:eff_tab_psitZgamma}
\end{table}

For the final channel of $\psi_t$, we consider the decay $\psi_t\to\gamma H$, which has a cross section of $\sigma(e^+e^-\to \psi_t\to \gamma H)=0.26$~fb at $\sqrt{s}=M_{\psi_t}$ according to the projection method after applying the basic photon cuts $p_{T}^\gamma>10~{\rm GeV}$ and $\vert\eta_\gamma\vert<3$. Unlike the previous channels, the significance of this process arises primarily from the resonance itself rather than from interference, since the SM production of $\gamma H$ proceeds only through loop processes and is suppressed. However, for Higgs detection through the $b\bar{b}$ decay, the dominant background becomes the continuous $\gamma b\bar{b}$ production.
For this channel, we require that the selected events contain at least one $b$-jet and pass the photon cuts mentioned above. To reconstruct the Higgs boson, we impose the mass window cut $120~{\rm GeV}< m_{b\bar{b}} < 128~{\rm GeV}$. Given the production of some boosted Higgs bosons from this process, the $b$-jets are often not resolved as separate jets by the detector, so we also include events with a single jet falling within this mass window, similar to the treatment for the previous channels. The most energetic photon is selected, and we further impose the cuts $|m_{\gamma H}-m_{\psi_t}| < 5$~GeV and $p_{T}^\gamma>85$~GeV.
The corresponding efficiencies are listed in Table~\ref{tab:eff_tab_psitHgamma}, yielding a significance of $0.6\,\sigma$ at the FCC-ee with $420~{\rm fb}^{-1}$. The possible measurement of this channel relies on the capability of future experiments to inclusively reconstruct the $\gamma H$ final state, which, if performed efficiently, would provide another way to discover $\psi_t$ and potentially probe the top Yukawa coupling.
\begin{table}[hbt!]
    \centering
    \renewcommand{\arraystretch}{1.25}
    \begin{tabular}{c|c|c|c}\hline\hline
        process & $\psi_t\to \gamma H$ & $(\gamma b \bar{b})_{\rm SM}$ & $(\gamma H)_{\rm SM}$ \\ \hline\hline
        cross section$\times$ BR [fb] & 0.1382 & 54.69 & 0.0466\\ \hline
        basic selections  & 69.63\% & 45.37\% & 70.38\%\\ \hline
        $H$ reconstruction  & 14.43\% & 2.97\% & 14.93\%\\ \hline
        hard photon cut  & 8.96\% & 0.334\% & 9.62\%\\ \hline\hline
        fiducial cross section [fb] & 0.0124 & 0.183 & 0.0045 \\ \hline\hline
    \end{tabular}
    \renewcommand{\arraystretch}{1}
    \caption{The cross sections and efficiencies for study of $\psi_t\to \gamma H$ with signal and SM background $\gamma b \bar{b}$, $\gamma H$. The branching ratio, BR, here for $\gamma H$ is the $H\to b\bar{b}$ branching ratio. To increase the statistics of $\gamma b \bar{b}$, we use $100~{\rm GeV}< m_{b\bar{b}} < 150$~GeV at generation level.}
    \label{tab:eff_tab_psitHgamma}
\end{table}

\subsection{$\chi_t$}\label{subsec:chi_t}

For the $P$-wave $\chi_t$ states, we consider both $\chi_{t0}$ and $\chi_{t1}$. Since they are both $P$-wave states, their production cross sections are both suppressed by $\vert R_P^\prime(0)\vert^2\propto \alpha_s^5$. For $\chi_{t0}$, we first consider its production via gluon fusion, whose parton-level cross section is given by
\begin{equation}
    \hat{\sigma}(gg\to\chi_{t0}) = \frac{12\pi^2\alpha_s^2}{M^3\hat{s}}\,\vert R_P^\prime(0)\,\vert^2\delta(\hat{s}-M^2) ~,
\end{equation}
with $M=M_{\chi_{t0}}$, which leads to the proton-level cross section $\sigma(pp\to\chi_{t0})=10.0$~fb at the 13-TeV LHC. This result is clearly prevailed by both $\sigma(gg\to\eta_{t})=8.8$~pb and the continuous $\sigma(gg\to t\bar{t})=833.9$~pb~\cite{CMS:2025kzt}. On the other hand, $\sigma(f\bar{f}\to\chi_{t0})\propto m_f^2$ is further suppressed by the small fermion masses both at the LHC ($f$ being the light quarks) and the lepton colliders ($f$ being the light charged leptons). Therefore, it is unlikely for $\chi_{t0}$ to be probed at the LHC and other future colliders.

For $\chi_{t1}$, again, it cannot be produced via gluon fusion due to the Landau-Yang theorem, but since its coupling to a light fermion pair is not suppressed by the fermion mass, one can consider its production at lepton colliders. One starts from the partial width
\begin{equation}
    \Gamma(\chi_{t1}\to f\bar{f}) = C_f\frac{96\,\alpha_{\rm EM}^2\,a_t^2}{(M^2-m_Z^2)^2c_w^4s_w^4}\sqrt{1-\frac{4m_f^2}{M^2}}\left[v_f^2\left(1+\frac{2m_f^2}{M^2}\right)+a_f^2\left(1-\frac{4m_f^2}{M^2}\right)\right]\vert R_P^\prime(0)\vert^2 ~,
\end{equation}
where $M=M_{\chi_{t1}}$, $C_f=1,3$ is the color factor for leptons and quarks, respectively, and
\begin{equation}
    v_i = \frac{I_{3L,i}}{2} - Q_is_w^2 ,\qquad  a_i = \frac{I_{3L,i}}{2} ~,
\end{equation}
with $Q_i$ being the electric charge of the fermion $i$ and $I_{3L}=+1/2~(-1/2)$ for up-type quarks and neutrinos (down-type quarks and charged leptons). For $f=e^-$, one can use the B-W formula to obtain $\sigma(e^+e^-\to\chi_{t1})=0.54$~fb. Naively estimating with the cross section of the continuous plus $\psi_t$ background $\sigma(e^+e^-\to t\bar{t}/\psi_t)\approx 1.5$~pb (see Figure~\ref{fig:xs_ee_psit+hardgamma}) at $\sqrt{s}=M_{\chi_{t1}}$, one gets a significance of only $\sim0.56\,\sigma$ with even $\mathcal{L}=1~{\rm ab}^{-1}$. Compared to the $S$-wave $\psi_t$ whose 
differential cross section $\dd\sigma/\dd\theta$ is uniformly distributed across the production angle $\theta$ of the constituent $t$ after the decay, that of the $P$-wave $\chi_{t1}$ depends non-trivially on $\theta$, and thus one could try to refine the analysis by analyzing the final-state kinematic distributions. However, we do not expect the significance to improve up to $5\,\sigma$ even after performing such analysis. On the other hand, $\sigma(q\bar{q}\to\chi_{t1})$ at the LHC still suffers from the large continuous $t\bar{t}$ background as in the case of $\chi_{t0}$. Therefore, we conclude that it is also unlikely to probe $\chi_{t1}$ at the LHC and other future colliders.

\section{Toponium as a probe for the real singlet extension}\label{sec:SSM}

In addition to probing the toponium states themselves, we further consider the prospect of probing BSM physics using the toponium measurement. Here we use the real singlet extension to the SM (SSM) as an example, though one can certainly generalize the framework to other BSM models. For simplicity, we consider the $\mathbb{Z}_2$-symmetric SSM, in which one introduces a $\mathbb{Z}_2$-odd real scalar $S$. Along with the SM Higgs doublet $\Phi$, the most general renormalizable scalar potential of the SSM is given by
\begin{align}
    V(\Phi, S)=-\mu^2\Phi^\dagger\Phi-m^2 S^2+\lambda_1 (\Phi^\dagger\Phi)^2 +\lambda_2 S^4+ \lambda_3S^2\Phi^\dagger\Phi ~,
\end{align}
where we follow the coupling normalization convention of Ref.~\cite{Robens:2015gla}. Parameterizing the fields as $\Phi=(0, \frac{h_1+v}{\sqrt{2}})^T$ and $S=\frac{h_2+x}{\sqrt{2}}$, where $v$ and $x$ denote the vacuum expectation values (VEVs) of the fields, respectively, one obtains the minimization conditions
\begin{equation}
    m^2 = \lambda_1v^2 + \frac{\lambda_3}{2}x^2 ,\qquad \mu^2 = \lambda_2x^2 + \frac{\lambda_3}{2}v^2 ~,
\end{equation}
which lead to the mass matrix in the $(h_1,h_2)^T$ basis as
\begin{equation}
    M^2 = \begin{pmatrix}
        2\,\lambda_1\,v^2 & \lambda_3\, v\,x \\[1ex]
        \lambda_3\, v\,x & 2\,\lambda_2\, x^2
    \end{pmatrix} ~.
\end{equation}
Denoting $\alpha$ as the mixing angle between $h_{1,2}$, one gets the mixing matrix
\begin{align}
    \begin{pmatrix}
    \phi\\
    H
    \end{pmatrix}=\begin{pmatrix}
    \cos\alpha & \sin\alpha\\
    -\sin\alpha & \cos\alpha
    \end{pmatrix}\begin{pmatrix}
    h_1\\
    h_2
    \end{pmatrix} ~,
\end{align}
where we identify $H$ with the 125-GeV Higgs boson observed at the LHC and $\phi$ with the exotic mass eigenstate with $m_\phi<m_H=125$~GeV. Under this mixing convention, the SM Higgs is recovered at the decoupling limit $\sin\alpha=-1$. Consequently, one can write down the top Yukawa interaction in the mass basis as
\begin{equation}\label{eq:Lag_SSM_Yukawa}
    \mathcal{L}_{\rm Yuk} = -\frac{m_t}{v} (\phi\cos\alpha - H\sin\alpha)\, \bar{t}t ~,
\end{equation}
which, in the position space, leads to the Yukawa potential
\begin{align}
    V_{\rm Yuk}(r)=-\frac{1}{4\pi r} \left[\left(\frac{m_t\cos\alpha}{v}\right)^2 e^{-m_\phi r} + \left(\frac{m_t\sin\alpha}{v}\right)^2 e^{-m_H r}\right] ~.
\end{align}
As a result, the static potential given in Eq.~\eqref{eq:V_static} will be modified by the inclusion of the $\phi$- and $H$-mediated Yukawa potentials, which allows us to use the $\eta_t$ production measurements to indirectly infer the existence of additional scalar fields coupling to the top quark. 
More specifically, after solving the Schr\"{o}dinger equation with the modified static potential $V_{\rm static}+V_{\rm Yuk}$, one can obtain the corresponding $E_{\rm bind}$ and $R_S(0)$. These can further be used to predict $\sigma(pp\to \eta_t)$, which can then be compared to the experimental measurements. For simplicity, we only use the results reported in Ref.~\cite{CMS:2025kzt} as a demonstration of the methodology, since, as we will show later, the current precision in $\eta_t$ measurements is yet high enough to be sensitive to the unexplored parameter space of the SSM.
Here, the Yukawa potential mediated by the 125-GeV Higgs boson is suppressed by a factor of $e^{-m_H a_t}\sim \mathcal{O}(10^{-3})$ and thus will not influence the properties of toponium, as already pointed out in Ref.~\cite{Strassler:1990nw}. On the contrary, the light scalar $\phi$ field could constructively contribute to the static potential (for a color-singlet toponium) and enhance the effective couplings within the toponium system with great significance. Consequently, the experimental measurements will exclude the parameter space where $\cos\alpha\gtrsim0$ (or where $\vert\sin\alpha\vert\lesssim1$), as we will show later.

\begin{figure}[th!]
    \centering
    \includegraphics[width=0.45\linewidth]{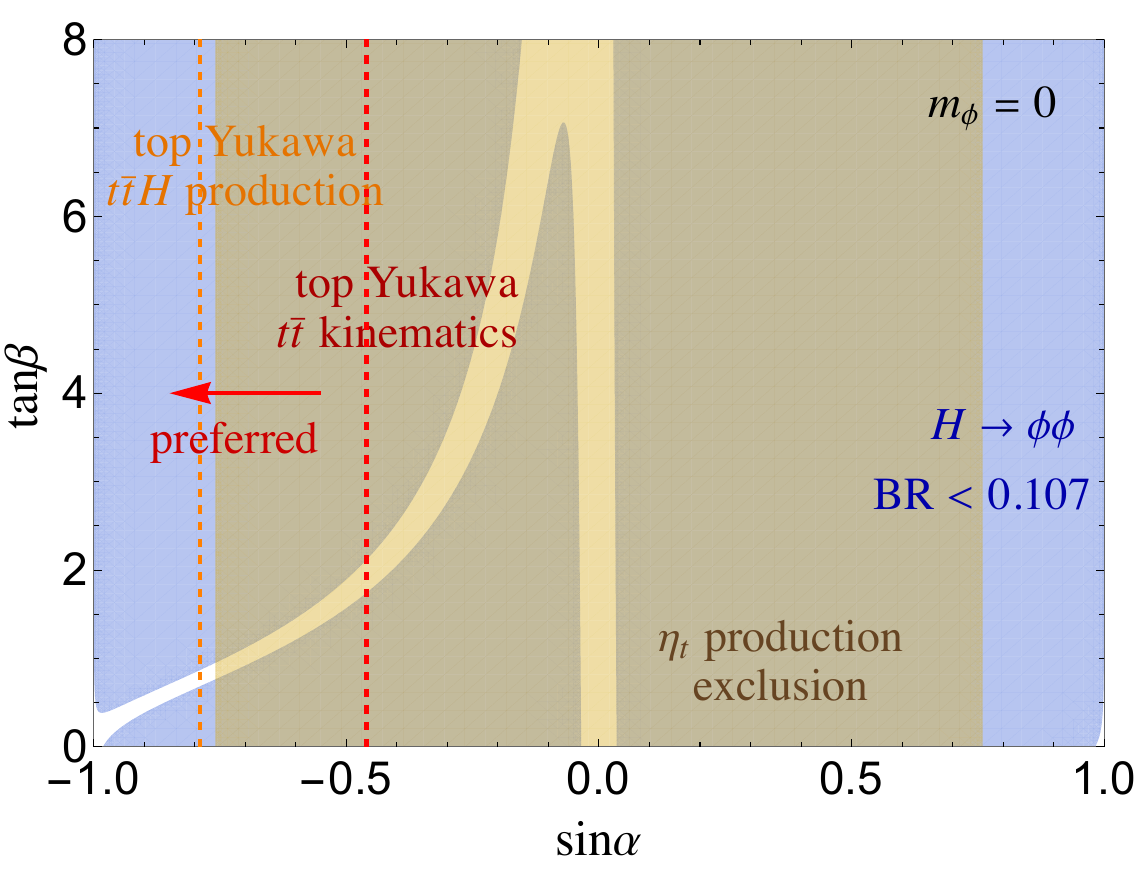}
    \includegraphics[width=0.45\linewidth]{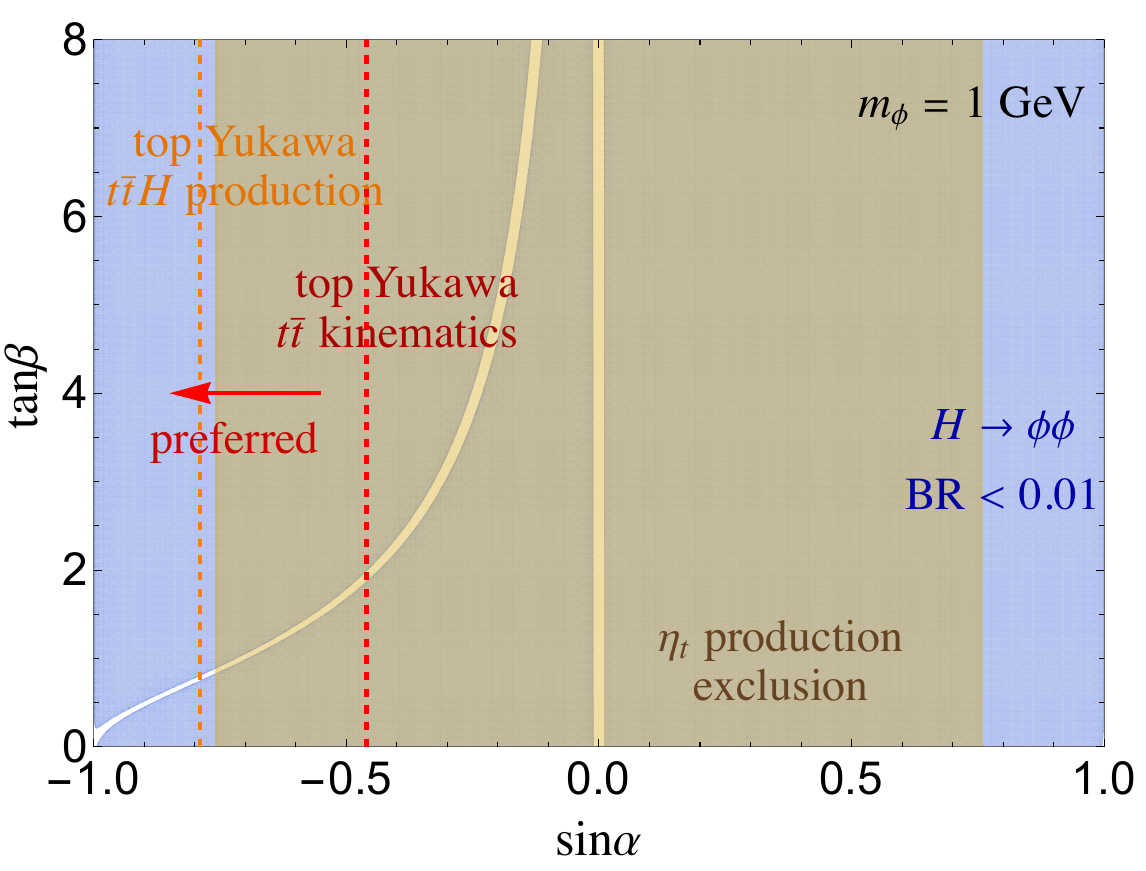}
    \includegraphics[width=0.45\linewidth]{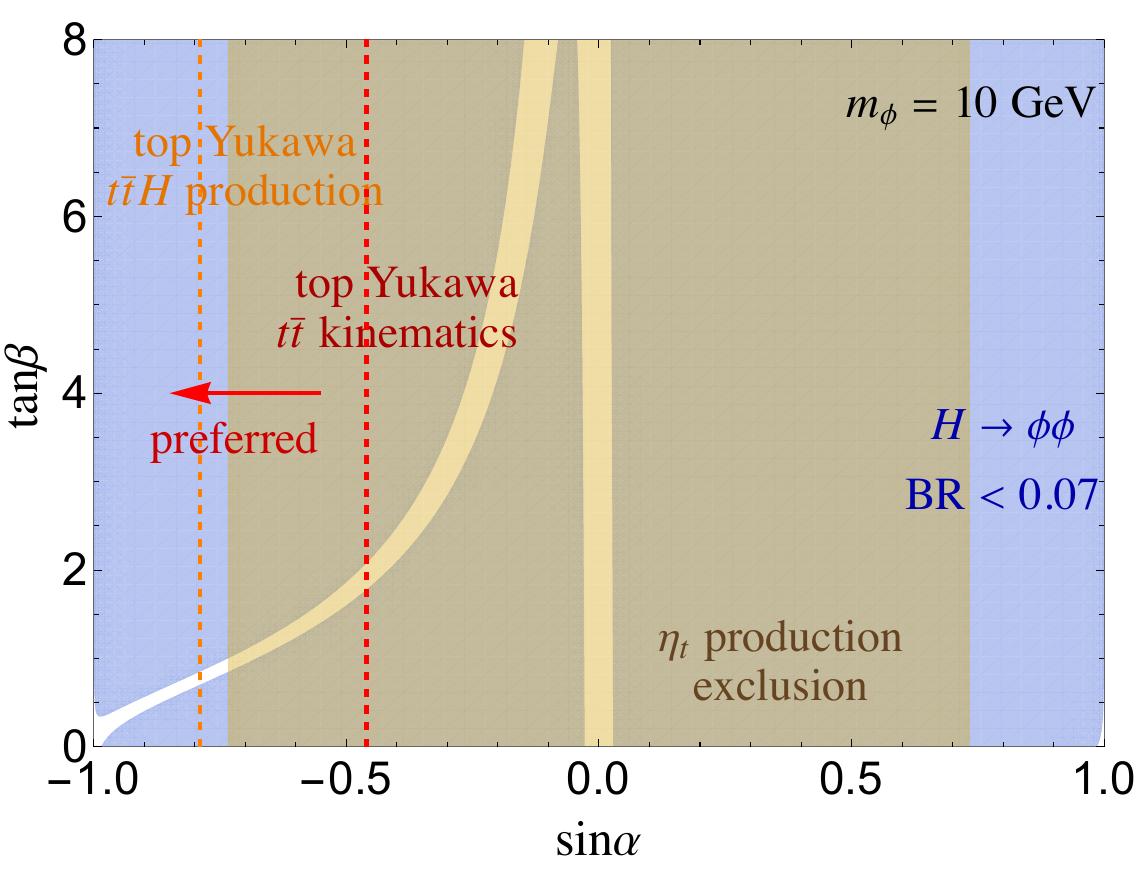}
    \includegraphics[width=0.45\linewidth]{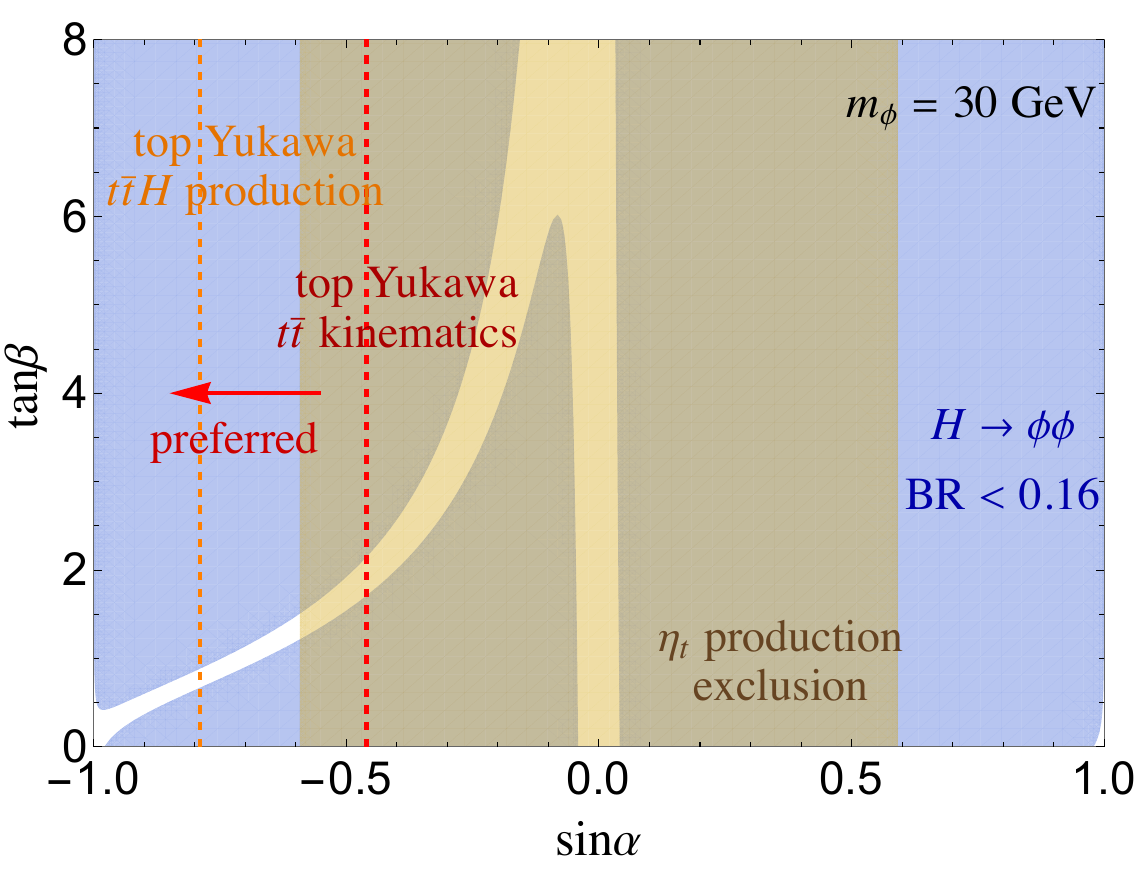}
    \caption{The 95\% CL constraint on ${\rm BR}(H\to\phi\phi)$ obtained from Refs.~\cite{Carena:2022yvx, ParticleDataGroup:2024cfk} (blue) and that imposed by the $\eta_t$ production measurement performed in Ref.~\cite{CMS:2025kzt} (brown) on the ($\sin\alpha$)--($\tan\beta$) plane for four benchmark masses $m_\phi=0,1,10,30$~GeV. We also show the preferred regions (red arrow) by the 125-GeV Higgs top Yukawa coupling measurements at 95\% CL summarized in Ref.~\cite{CMS:2024irj} with two dashed lines: the red one is measured through the $t\bar{t}$ kinematics~\cite{CMS:2020djy} and the orange one through the $t\bar{t}H$ production cross section~\cite{ATLAS:2022vkf, CMS:2022dwd}.
    }
    \label{fig:z2sssm_cosa_tanb}
\end{figure}

On the other hand, the current Higgs exotic decay can also be used to constrain this model. Denoting $\tan\beta\equiv v/x$, the width of the decay channel $H\to\phi\phi$ for $m_\phi < m_H/2$ is given by~\cite{Robens:2015gla}
\begin{align}
    \Gamma(H\to\phi\phi)=\frac{1}{8\pi m_H}\sqrt{1-\frac{4m_\phi^2}{m_H^2}}\left[\frac{\sin2\alpha}{2v}(\tan\beta\sin\alpha+\cos\alpha)\left(m_\phi^2+\frac{m_H^2}{2}\right)\right]^2 ~.
\end{align}
We show the 95\% CL constraint on ${\rm BR}(H\to\phi\phi)$ on the ($\sin\alpha$)--($\tan\beta$) plane obtained from Ref.~\cite{Carena:2022yvx, ParticleDataGroup:2024cfk} for four benchmark masses $m_\phi=0,1,10,30$~GeV in Figure~\ref{fig:z2sssm_cosa_tanb}.  
The 95\% CL constraints with regard to the Yukawa potential variation derived from the $\eta_t$ measurement are also presented in Figure~\ref{fig:z2sssm_cosa_tanb}, from which one can see that for all four benchmark masses, the allowed parameter space by the ${\rm BR}(H\to\phi\phi)$ constraint, which roughly follows the $\tan\beta\sin\alpha+\cos\alpha=0$ contour that gives ${\rm BR}(H\to\phi\phi)=0$, is further ruled out by the $\eta_t$ production measurement down to the region around $(\sin\alpha,\tan\beta)=(-1,0)$. Since the top Yukawa coupling of the 125-GeV Higgs boson is also modified to $y_t=-m_t\sin\alpha/v$ [see Eq.~\eqref{eq:Lag_SSM_Yukawa}], we show in Figure~\ref{fig:z2sssm_cosa_tanb} the preferred ranges of $\sin\alpha$ by two types of top Yukawa coupling measurements (see Ref.~\cite{CMS:2024irj} for a review): the direct measurement of $t\bar{t}H$ production cross section gives $y_t=0.95\substack{+0.07\\-0.08}$~\cite{ATLAS:2022vkf, CMS:2022dwd}, while the measurement of the loop-induced correction to the kinematic distribution of the $t\bar{t}$ final state, which mainly comes from the top Yukawa coupling between $t$ and $\bar{t}$, gives $y_t=1.16\substack{+0.24\\-0.35}$~\cite{CMS:2020djy}. It can be seen that the exclusion by the $\eta_t$ measurement has not yet reached the current limits imposed by the top Yukawa coupling measurements. However, as future measurements of $\eta_t$ are expected to improve further, they could potentially exclude a larger region of the parameter space. Finally, while we only present the study on the SSM as an example, one can certainly generalize this framework to other BSM models that afford a more sophisticated top Yukawa coupling structure, for which the $\eta_t$ measurement may impose a more non-trivial constraint.

\section{Discussion and conclusions}\label{sec:conclusions}

We remark again that, as we have discussed and demonstrated in Section~\ref{subsec:eta_t} through Figure~\ref{fig:BindE_XS_pp_etat}, the theoretical predictions of toponium properties performed in Refs.~\cite{CMS:2025kzt,ATLAS:2026dbe} still suffer from the uncertainty arising from $M_{\eta_t}$. Therefore, it is important to revisit the analysis to pin down the production cross section $\sigma(pp\to\eta_t)$ in a more rigorous manner, which, along with the measured toponium mass, could provide a measure for the top quark mass and width (see Ref.~\cite{CMS:2024irj} for a review of the existing top quark mass measurements). Since $\eta_t$ and $\psi_t$ have approximately the same mass, a precise measurement of the $\eta_t$ mass at the HL-LHC can help determine the center-of-mass energy around the $t\bar{t}$ threshold to be scanned at FCC-ee (see Ref.~\cite{Beneke:2015kwa} for the N$^3$LO QCD calculation and Ref.~\cite{Defranchis:2025auz} for projected measurements of the top quark mass and width near the $t\bar{t}$ threshold region at a future lepton collider).  

Besides the projection and Green function methods, one can also use the ``bound state effective field theory (BSEFT)'' to study toponium decays and production, such as introduced in Ref.~\cite{Fu:2025zxb}. However, since the effective couplings in the BSEFT are derived under the assumption that all participant particles are on-shell, it can only be used to study the processes in which the fields interacting with the toponium are all external. For example, the $\psi_tb\bar{b}$ vertex derived from the BSEFT is given by $\bar{b}\gamma_\mu(g^V_{\psi_t b}-g^A_{\psi_t b}\gamma^5)b \psi_t^\mu$, where
\begin{align}
    g^V_{\psi_t b}=&~\sqrt{\frac{(C_F\alpha_s)^3\pi}{3}}\frac{\alpha_{\rm EM}}{12}\left\{\frac{m_t^2}{s_w^2}\left[\frac{4(8c_w^2+1)s_w^2-9}{c_w^2(Q^2-m_Z^2)}+\frac{3(2-K^2/m_W^2)}{K^2-m_W^2}\right]-8\frac{4m_t^2}{Q^2}\right\} ~,\label{eqn:gVPsitBB}\\
    g^A_{\psi_t b}=&~-\sqrt{\frac{(C_F\alpha_s)^3\pi}{3}}\frac{\alpha_{\rm EM}\,m_t^2}{4s_w^2}\left[\frac{3-8s_w^2}{c_w^2(Q^2-m_Z^2)}-\frac{2-K^2/m_W^2}{K^2-m_W^2}\right] ~,\label{eqn:gAPsitBB}
\end{align}
with $k_{1,2}$ denoting the outgoing momenta of $b$ and $\bar{b}$, respectively, and, for simplicity, $Q^2\equiv (k_1+k_2)^2=M_{\psi_t}^2$ and $K^2\equiv(k_1-k_2)^2/4=-M_{\psi_t}^2/4$ in the $m_b\to0$ limit. However, in the general case where $b$ and $\bar{b}$ are not necessarily on-shell, not only will $Q^2$ and $K^2$ depend on the exact kinematics of the scattering process, but there will also be an additional term $\propto K^\mu$ that can only be captured by the projection method. We have checked that this discrepancy will lead to a huge difference in the predicted production cross section of the $pp\to \psi_tb\bar{b}$ process, which is also the case for other interactions and corresponding production mechanisms. Therefore, one should be careful about using the BSEFT to study toponium phenomenology.

In addition to the physics discussed in the main text, one might also consider probing CP-violation in the top Yukawa coupling via the annihilation decay modes of toponia that involve $H$. For instance, in the presence of CP-violation, the $\psi_t\to\gamma H$ decay width [see Eq.~\eqref{eqn:psit_gammaH} for the corresponding formula] will scale from $\vert y_t^2\vert\to\vert y_t\vert^2 + \vert y_t^\prime\vert^2$, $y_t$ and $y_t^\prime$ denoting the possible CP-even and CP-odd components of top Yukawa coupling, respectively, while some originally forbidden decay modes such as $\eta_t\to HH$ will be turned on. Although these effects are difficult to be measured and tightly constrained by the experimental measurements to date, it still remains an interesting question whether one can use these toponium decays to probe certain BSM physics that introduces CP-violation.

Another important implication of toponium physics is the effect of $\psi_t$ on the electroweak precision tests, which has been studied in Ref.~\cite{Kniehl:1992dx}. Denoting the Fermi constant as $G_F$ and using the Green function method, one can show that~\footnote{We are not aware that the public \texttt{HEPfit} package~\cite{DeBlas:2019ehy} for global electroweak precision observable fits has implemented this correction.}
\begin{align}
    \Delta \rho(\mu) &= \frac{G_F}{8\sqrt{2}}N_cC_F\frac{\alpha_s(\mu)}{\pi}m_tE_{\rm bind} ~, \\
    \Delta r(\mu) &= -\frac{c_w^2}{s_w^2}\Delta\rho(\mu)\left[
        1-\left(1-\frac{8}{3}s_w^2\right)^2\frac{m_Z^2}{(2m_t)^2-m_Z^2}
    \right] - \frac{N_c\,C_F\,\alpha_{\rm EM}\,\alpha_s(\mu)E_{\rm bind}}{9\,m_t} ~,
\end{align}
which, after plugged into
\begin{equation}
    m_W^2 = \frac{m_Z^2}{2}\left\{
        1 + \sqrt{
            1 - \frac{4\pi\,\alpha_{\rm EM}}{\sqrt{2}\,G_F\,m_Z^2}\left[1+\Delta r(m_W)\right]
        }
    \right\} ~,
\end{equation}
along with $E_{\rm bind}=-2.725$~GeV, gives $\Delta m_W\approx -4.0$~MeV, which is comparable to the state-of-the-art measurement uncertainty ($\approx10$~MeV) performed by CMS~\cite{CMS:2024lrd}. While of course one needs to further consider other sources of corrections, as one of the main objectives of FCC-ee is an even more precise test on the electroweak theory, a more comprehensive understanding of threshold $t\bar{t}$ and $\psi_t$ effects is crucial to the interpretation of future experiments.

In conclusion, we have calculated some theoretical properties of toponia, including their masses, decays, and production at colliders. To study their annihilation decay modes and production mechanisms, we have adopted both projection and Green function methods and found that they both gave comparable results to the CMS measurement of the $pp\to\eta_t$ production cross section at the 13-TeV LHC. Using the projection method, we have studied the sensitivity of the four lowest-lying toponium states at the current LHC and future HL-LHC and FCC-ee, where we have found that $\eta_t$ is likely to be probed via the associated production of $\eta_tb\bar{b}$ at the hadron colliders and $\psi_t$ at the lepton colliders, while the other two $P$-wave states, $\chi_{t0}$ and $\chi_{t1}$, seem unlikely to be probed in the near future. Finally, we have also used the SSM as an example to show how one can constrain the parameter space of BSM models with the $\eta_t$ measurement.

\vspace{0.5cm}
\subsubsection*{Acknowledgments}
We thank Kaoru Hagiwara, Cheng-Ping Shen and Yu-Jie Zhang for useful discussions. This work is supported by the U.S. Department of Energy under the contract DE-SC-0017647 and DEAC02-06CH11357 at Argonne National Laboratory. TKC is also supported by the Ministry of Education, Taiwan, under the Government Scholarship to Study Abroad.

\appendix
\section{Partial width formulas for the annihilation decay modes}\label{subapp:toponium_decay_width}

In this appendix, we summarize the formulas for the annihilation decay widths of the toponium states adapted from Refs.~\cite{Kuhn:1980gw,Kuhn:1987ty,Barger:1987xg}. We denote $x_X\equiv m_X^2/M^2$, where $M$ is the mass of the corresponding toponium, and define the K\"{a}ll\'{e}n function $\lambda(a,b,c)\equiv\sqrt{a^2+b^2+c^2-2ab-2bc-2ca}$. For $\psi_t\to b\bar{b}$ and $\chi_{t1}\to b\bar{b}$, there is an additional contribution from the $W$-exchange diagrams compared to the other $f\bar{f}$ decay modes. To turn on this specifically for $f=b$, we define $\delta_{fb}=1$ if $f=b$ and $\delta_{fb}=0$ if otherwise. For simplicity, we set $m_f=0$ in the following expressions.

\hfill \\
\noindent $\bullet~\eta_t~(0^{-+})$:

\begin{align}
    \Gamma(\eta_t\to W^+ W^-)=&\frac{3\alpha_{\rm EM}^2}{2M^2s_w^4(1-4x_W)^2}\lambda^{3/2}(1,x_W,x_W)|R_S(0)|^2 ~, \\
    \Gamma(\eta_t\to ZZ)=~&\frac{\alpha_{\rm EM}^2(32s_w^4-24s_w^2+9)^2}{432M^2s_w^4c_w^4(1-2x_Z)^2}\lambda^{3/2}(1,x_Z,x_Z)|R_S(0)|^2 ~ ,\\
    \Gamma(\eta_t\to ZH)=~&\frac{3\alpha_{\rm EM}^2}{64M^2s_w^4c_w^4x_Z^2}\lambda^{3/2}(1,x_H,x_Z)|R_S(0)|^2 ~, \\
    \Gamma(\eta_t\to\gamma\gamma) =~&\frac{64\alpha_{\rm EM}^2}{27M^2}|R_S(0)|^2 ~, \\
    \Gamma(\eta_t\to \gamma Z)=~&\frac{2\alpha_{\rm EM}^2(3-8s_w^2)^2}{27M^2s_w^2c_w^2}\lambda^{1/2}(1,0,x_Z)|R_S(0)|^2 ~, \\
    \Gamma(\eta_t\to gg) =~&\frac{8\alpha_s^2}{3M^2}|R_S(0)|^2 ~.
\end{align}

\hfill \\
\noindent $\bullet~\psi_t~(1^{--})$:

\begin{align}
    \Gamma(\psi_t\to f\bar{f})=~&\frac{4C_f\alpha_{\rm EM}^2}{M^2}\Bigg\{\left[\frac{2}{3}Q_f+\frac{v_f(1-\frac{8}{3}s_w^2)}{4s_w^2c_w^2(1-x_Z)}-\frac{\delta_{fb}(1+8x_W)}{48s_w^2(1+4x_W)x_W}\right]^2\nonumber\\
    &+\left[\frac{a_f(1-\frac{8}{3}s_w^2)}{4s_w^2c_w^2(1-x_Z)}-\frac{\delta_{fb}(8x_W+1)}{48s_w^2(1+4x_W)x_W}\right]^2\Bigg\}|R_S(0)|^2 ~,\\
    \Gamma(\psi_t\to W^+ W^-)=~&\frac{\alpha_{\rm EM}^2}{64M^2x_W^2s_w^4}\lambda^{3/2}(1,x_W,x_W)\Bigg[\frac{1+20x_W+12x_W^2}{(1-x_Z)^2}\left(1-\frac{16}{3}s_w^2 x_Z+\frac{64}{9}s_w^4 x_Z^2\right)\nonumber\\
    &-\frac{8x_W(5+6x_W)(1-\frac{8}{3}s_w^2x_Z)}{(1-4x_W)(1-x_Z)}+\frac{16x_W(2-x_W)}{(1-4x_W)^2}\Bigg]|R_S(0)|^2 ~,\\
    \Gamma(\psi_t\to ZZ)=~&\frac{\alpha_{\rm EM}^2(1-\frac{8}{3}s_w^2)^2}{32M^2s_w^4c_w^4(1-2x_Z)^2x_Z}\lambda^{5/2}(1,x_Z,x_Z)|R_S(0)|^2 ~,\\
    \Gamma(\psi_t\to \gamma H)=~&\frac{4\alpha_{\rm EM}y_t^2}{9\pi M^2}(1-x_H)|R_S(0)|^2=\frac{2\alpha_{\rm EM}}{9\pi v_{\text{EW}}^2}(1-x_H)|R_S(0)|^2 ~,\\
    \Gamma(\psi_t\to ZH)=~&\frac{\alpha_{\rm EM}^2(1-\frac{8}{3}s_w^2)^2}{32M^2s_w^4c_w^4}\lambda^{1/2}(1,x_H,x_Z)\\
    &\times \frac{[(1-x_Z)^2-x_H(1-3x_Z)]^2+\frac{1}{2}x_Z[(1-x_Z)^2+x_H(2-x_H)]^2}{x_Z(1-x_Z)^2(1-x_Z-x_H)^2}|R_S(0)|^2 ~,\nonumber\\
    \Gamma(\psi_t\to \gamma Z)=~&\frac{2\alpha_{\rm EM}^2(1-x_Z^2)}{9M^2s_w^2c_w^2x_Z}|R_S(0)|^2 ~, \\
    \Gamma(\psi_t\to 3g) =~& \frac{40(\pi^2-9)\alpha_s^3}{81\pi M^2}|R_S(0)|^2 ~.\label{eqn:psit_gammaH}
\end{align}

\hfill \\
\noindent $\bullet~\chi_{t0}~(0^{++})$:

\begin{align}
    \Gamma(\chi_{t0}\to W^+ W^-)=~&\frac{3\alpha_{\rm EM}^2}{4M^4s_w^4  x_W^2}\lambda^{1/2}(1,x_W,x_W)\Bigg\{\left[\frac{2(1-x_W)}{1-4x_W}-\frac{3(1-2x_W)}{2(1-x_H)}\right]^2\nonumber\\
    &+2x_W^2\left[\frac{2}{1-4x_W}+\frac{3}{1-x_H}\right]^2\Bigg\}|R_P'(0)|^2 ~, \\
    \Gamma(\chi_{t0}\to ZZ)=~&\frac{96\alpha_{\rm EM}^2}{M^4 s_w^4c_w^4x_Z^2}\lambda^{1/2}(1,x_Z,x_Z)\Bigg\{\left[\frac{(1-\frac{8}{3}s_w^2)^2x_Z^2}{8(1-2x_Z)^2}-\frac{1}{16}+\frac{3}{32}\frac{1-2x_Z}{1-x_H}\right]^2\\
    &+\frac{1}{2}x_Z^2\left[\frac{-(1-\frac{8}{3}s_w^2)^2(3-8x_Z)+(3-4x_Z)}{16(1-2x_Z)}-\frac{3}{8(1-x_H)}\right]^2\Bigg\}|R_P'(0)|^2 ~,\nonumber\\
    \Gamma(\chi_{t0}\to HH)=~&\frac{3\alpha_{\rm EM}^2}{32M^4s_w^4x_W^2}\lambda^{1/2}(1,x_H,x_H)\left[\frac{5-8x_H}{(1-2x_H)^2}-\frac{9x_H}{1-x_H}\right]^2|R_P'(0)|^2 ~,\\
    \Gamma(\chi_{t0}\to\gamma\gamma) =~&\frac{256\alpha_{\rm EM}^2}{3M^4}|R_P'(0)|^2 ~, \\
    \Gamma(\chi_{t0}\to Z\gamma)=~&\frac{24\alpha_{\rm EM}^2}{M^4s_w^2c_w^2}\left(1-\frac{8}{3}s_w^2\right)^2\frac{(1-\frac{1}{3}x_Z)^2}{1-x_Z}|R_P'(0)|^2 ~, \\
    \Gamma(\chi_{t0}\to gg) =~&\frac{96\alpha_s^2}{M^4}|R_P'(0)|^2 ~.
\end{align}

\hfill \\
\noindent $\bullet~\chi_{t1}~(1^{++})$:

\begin{align}
    \Gamma(\chi_{t1}\to f\overline{f})=~&\frac{96C_f\alpha_{\rm EM}^2}{M^4}\Bigg\{\left[\frac{a_f}{4s_w^2c_w^2(1-x_Z)}-\frac{\delta_{fb}(-1+8x_W)}{48s_w^2(1+4x_W)x_W}\right]^2\nonumber\\
    &+\left[\frac{v_f}{4s_w^2c_w^2(1-x_Z)}-\frac{\delta_{fb}(-1+8x_W)}{48s_w^2(1+4x_W)x_W}\right]^2\Bigg\}|R_P'(0)|^2 ~,\\
    \Gamma(\chi_{t1}\to W^+W^-)=~&\frac{3\alpha_{\rm EM}^2}{2M^4s_w^4x_W(1-4x_W)^2}\lambda^{3/2}(1,x_W,x_W)\bigg[\frac{64x_W^2}{1-4x_W}+4\left(\frac{x_Z-4x_W}{1-x_Z}\right)^2\nonumber\\
    &+\frac{(1+4x_W-2x_Z)^2(1+4x_W+12x_W^2)}{4x_W(1-x_Z)^2}\bigg]|R_P'(0)|^2 ~,\\
    \Gamma(\chi_{t1}\to ZZ)=~&\frac{3\alpha_{\rm EM}^2}{16M^4s_w^4c_w^4x_Z(1-2x_Z)^2}\lambda^{5/2}(1,x_Z,x_Z)\left[1-\frac{2(1-\frac{8}{3}s_w^2)^2x_Z}{1-2x_Z}\right]^2|R_P'(0)|^2 ~,\\
    \Gamma(\chi_{t1}\to ZH)=~&\frac{3\alpha_{\rm EM}^2}{8s_w^4c_w^4M^4 x_Z^2(1-x_Z-x_H)^2}\lambda^{1/2}(1,x_Z,x_H)  \\
    &\times \Bigg\{(1+x_Z-x_H)^2\left(1-x_Z+\frac{x_H x_Z}{1-x_Z}\right)^2 +\frac{1}{2}x_Z \nonumber\\
    &\times\left[3-3x_Z+x_H+4x_Hx_Z\left(\frac{1}{1-x_Z}+\frac{1}{1-x_Z-x_H}\right)\right]^2\Bigg\}|R_P'(0)|^2 ~,\nonumber\\
    \Gamma(\chi_{t1}\to \gamma Z)=~&\frac{16\alpha_{\rm EM}^2}{3M^4s_w^2c_w^2}\left(1-\frac{8}{3}s_w^2\right)^2\frac{x_Z(1+x_Z)}{1-x_Z}|R_P'(0)|^2 ~.
\end{align}

We also list here the cross section formulas for $e^+e^-\to b\bar{b}$ and $\gamma Z$, including the interference terms with the background processes.

\begin{align}
    &\sigma(e^+e^-\to b\bar{b})=\\
    &\frac{\sqrt{s-4m_b^2}}{144\pi s^{5/2}s_w^4c_w^4 (s-m_Z^2)^2\bigl[M_{\psi_t}^2(\Gamma_{\psi_t}^2-2s)+M_{\psi_t}^4+s^2\bigr]}\nonumber\\
    &\times\biggl\{ -3\pi\alpha_{\rm EM} s c_w^6s_w^2\Bigl[6s(g_{\psi_t \ell}^A-g_{\psi_t \ell}^V)(M_{\psi_t}^2-s)(m_Z^2-s)\bigl(g_{\psi_t b}^V(2m_b^2+s)-g_{\psi_t b}^A(s-4m_b^2)\bigr)\nonumber\\
    &+4\pi\alpha_{\rm EM}(2m_b^2+s)(2m_Z^2-s)\bigl(M_{\psi_t}^2(\Gamma_{\psi_t}^2-2s)+M_{\psi_t}^4+s^2\bigr)\Bigr]\nonumber\\
    &-\pi  \alpha_{\rm EM}  s c_w^2 s_w^6 \Bigl[6 s (g_{\psi_t \ell}^A+3 g_{\psi_t \ell}^V) (M_{\psi_t }^2-s) (s-m_Z^2) \bigl(3 g_{\psi_t b}^A (s-4 m_b^2)+g_{\psi_t b}^V (2 m_b^2+s)\bigr)\nonumber\\
    &+4 \pi  \alpha_{\rm EM}  \bigl(4 m_b^2 (3 m_Z^2+8 s)+s (6 m_Z^2-11 s)\bigr) \bigl(M_{\psi_t }^2 (\Gamma_{\psi_t}^2-2 s)+M_{\psi_t }^4+s^2\bigr)\Bigr]\nonumber\\
    &+2 c_w^2 s_w^4 \biggl[18 s^2 ((g_{\psi_t \ell}^A)^2+(g_{\psi_t \ell}^V)^2) \left(s-m_Z^2\right)^2 \bigl((g_{\psi_t b}^A)^2 (s-4 m_b^2)+(g_{\psi_t b}^V)^2 (2 m_b^2+s)\bigr)\nonumber\\
    &-6 \pi  \alpha_{\rm EM}  s \Bigl[(M_{\psi_t }^2-s) (m_Z^2-s) (g_{\psi_t b}^V (2 m_b^2+s) \bigl(s g_{\psi_t \ell}^A+g_{\psi_t \ell}^V (8 m_Z^2-3 s)\bigr)-3 s g_{\psi_t b}^A (s-4 m_b^2) (g_{\psi_t \ell}^A+g_{\psi_t \ell}^V)\Bigr]\nonumber\\
    &+\pi ^2 \alpha_{\rm EM} ^2 \Bigl[m_b^2 (-48 s m_Z^2+64 m_Z^4-5 s^2)+s (-24 s m_Z^2+32 m_Z^4+11 s^2)\Bigr] \bigl(M_{\psi_t }^2 (\Gamma_{\psi_t }^2-2 s)+M_{\psi_t }^4+s^2\bigr)\biggl]\nonumber\\
    &- \pi ^2 \alpha_{\rm EM} ^2 s^2 \left[5s_w^8 (17 m_b^2-5 s)+9c_w^8 (m_b^2-s)\right] \bigl[M_{\psi_t }^2 (\Gamma_{\psi_t }^2-2 s)+M_{\psi_t }^4+s^2\bigr]\biggr\}\,,\nonumber
\end{align}
where $g_{\psi_t b}^{A,V}$ is defined in Eq.~\eqref{eqn:gAPsitBB_onshell} and
\begin{align}
    g_{\psi_t \ell}^{A}=&~-\sqrt{\frac{(C_F\alpha_s)^3\pi}{3}}\frac{\alpha_{\rm EM}\,m_t^2(3-8s_w^2)}{4s_w^2c_w^2(4m_t^2-m_Z^2)}\,,\nonumber\\
    g_{\psi_t \ell}^{V}=&~\sqrt{\frac{(C_F\alpha_s)^3\pi}{3}}\frac{\alpha_{\rm EM}[8m_W^2s_w^2-3m_t^2(1+4s_w^2)]}{4s_w^2c_w^2(4m_t^2-m_Z^2)}\,,
\end{align}
are the couplings of $\psi_t$ to leptons defined in $\bar{\ell}\gamma_\mu(g^V_{\psi_t \ell}-g^A_{\psi_t \ell}\gamma^5)\ell \psi_t^\mu$.

\begin{align}
    &\sigma(e^+e^-\to \gamma Z)=\\
    &\frac{-1}{384 \pi  s^2 s_w^2 c_w^2 m_Z^2 \left(s-m_Z^2\right) \bigl[(s-M_{\psi_t }^2)^2+M_{\psi_t }^2 \Gamma_{\psi_t }^2\bigr]}\int_{t_{\rm min}}^{t_{\rm max}}\dd t\nonumber\\
    &\biggl\{-8 c_w^2 s_w^2 t g_{\psi_t ZA}^2 \left[(g_{\psi_t \ell}^A)^2+(g_{\psi_t \ell}^V)^2\right] \left(s-m_Z^2\right) \Bigl[-m_Z^2 \left(6 s^2+9 s t+4 t^2\right)+3 m_Z^4 (s+2 t)+s \left(3 s^2+3 s t+2 t^2\right)\Bigr]\nonumber\\
    &+96 \pi  \alpha_{\rm EM}  \left(4 s_w^2-1\right) s_w c_w t g_{\psi_t ZA} m_Z^2 (g_{\psi_t \ell}^A+g_{\psi_t \ell}^V) \left(s-M_{\psi_t }^2\right) \left(s^2-m_Z^4\right)\nonumber\\
    &+96 \pi ^2 \alpha_{\rm EM} ^2 \left(8 s_w^4-4 s_w^2+1\right) m_Z^2 \bigl[(s-M_{\psi_t }^2)^2+M_{\psi_t }^2 \Gamma_{\psi_t }^2\bigr] \left[(m_Z^4+s^2) \log (\frac{t}{-m_Z^2+s+t})+2 t( s- m_Z^2)\right]\biggr\}\,,\nonumber
\end{align}
where $g_{\psi_t ZA}$ is defined in Eq.~\eqref{eqn:gPsitZA_onshell} and $t_{\rm min}$ and $t_{\rm max}$ are determined by the kinematic cuts for IR regularization on the photon. In terms of the maximum scattering angle $\theta_m$ with respect to the collision axis, $t_{\rm min/max}=-(s-m_Z^2)(1\mp \cos\theta_m)$.

\section{Green function method for $P$-wave states}\label{subapp:GF:diff}

\begin{figure}[th!]
    \centering
    \includegraphics[width=0.6\linewidth]{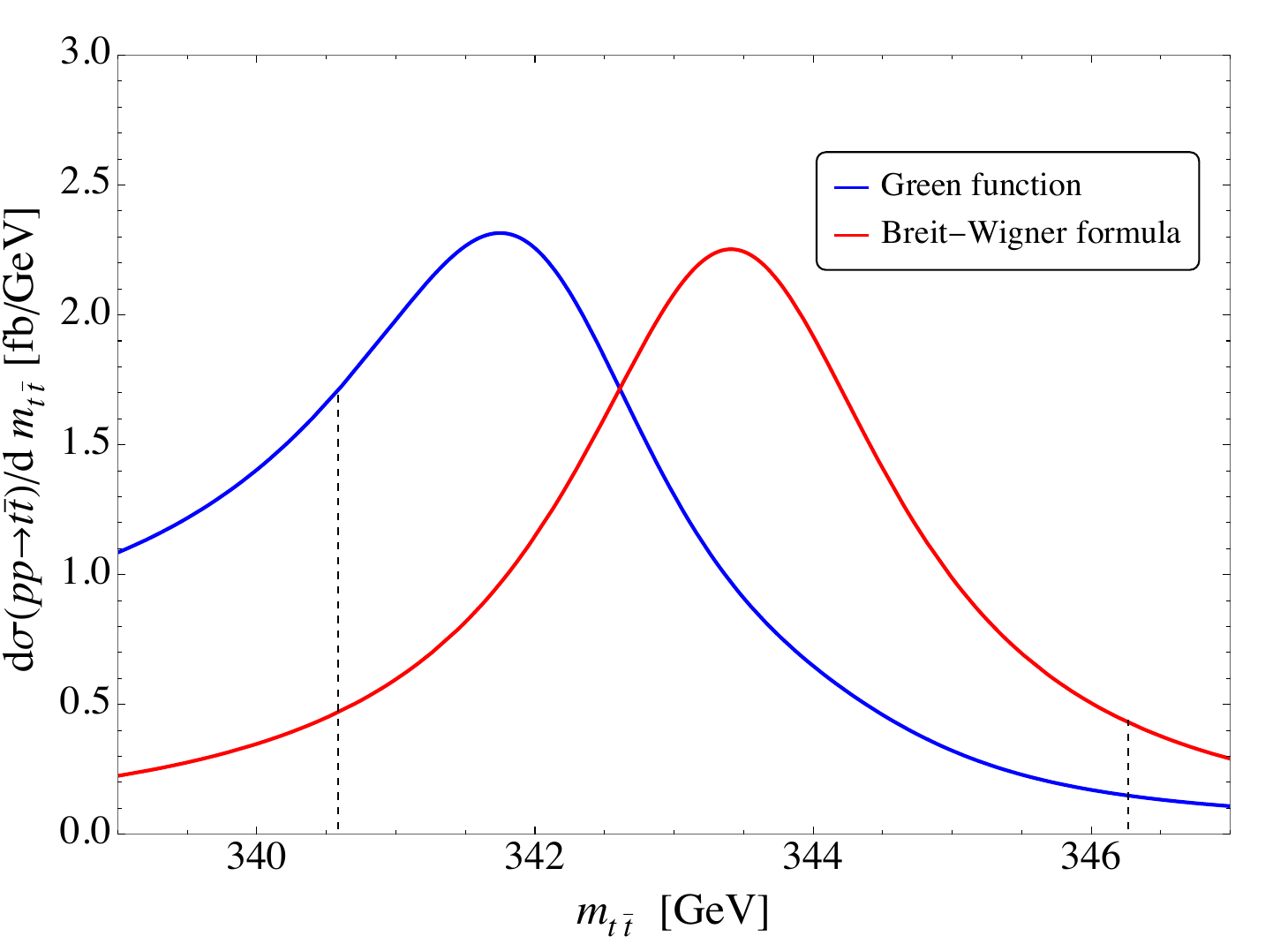}
    \caption{The $\dd\sigma(pp\to\chi_{t0})/\dd m_{t\bar{t}}$ distributions at the 13-TeV LHC calculated from the leading-order Green function method and B-W formula. The two dashed lines represent $m_{t\bar{t}}=M_{\chi_{t0}}\pm\Gamma$ with $M_{\chi_{t0}}=343.43~{\rm GeV}$ and $\Gamma=2.84~{\rm GeV}$, respectively, between which we identify the $t\bar{t}$ system to be on ``resonance''.
    }
    \label{fig:xs_pp_chi0t_GreenFunc}
\end{figure}

In this appendix, we compare the Green function and projection methods for $P$-wave toponium states. Following Ref.~\cite{deLima:2022joz}, one can write down
\begin{align}
    \sigma(gg\to t\bar{t}\to bW^+\bar{b}W^-)_{P}=\sigma(gg\to t\bar{t}\to bW^+\bar{b}W^-)_{P,\text{tree}}\frac{\Im[\nabla^2 G(r,E)]_{r=0}}{\Im[\nabla^2 G_0(r,E)]_{r=0}} ~,
\end{align}
where, approximating the static potential with $V_{\rm Coul}$ and using Eq.~\eqref{eq:green_analytic}, one has
\begin{align}
    \mathcal{S}_P(E)\equiv\frac{\Im[\nabla^2G(r,E)]_{r=0}}{\Im[\nabla^2G_0(r,E)]_{r=0}}=\frac{\Im\left[-a^2\log\left(-2i\sqrt{z'}\right)-a^2\,\psi^{(0)}\left(1-i\frac{a}{2\sqrt{z'}}\right)+ia\sqrt{z'}+z'\right]}{\Im\left[z'\right]} ~.
\end{align}
Using the $\chi_{t0}$ state for example, its parton-level production cross section $\hat{\sigma}(gg\to t\bar{t})_{^3P_0,\text{tree}}$ near the threshold is given by
\begin{align}
    \hat{\sigma}(gg\to t\bar{t})_{^3P_0,\text{tree}}=\frac{\pi\,\alpha_s^2\,(24m_t^2+\hat{s})}{144\,\hat{s}\,(2m_t+\sqrt{\hat{s}})^2}\left(1-\frac{4m_t^2}{\hat{s}}\right)^{\frac{3}{2}} ~,
\end{align}
where we have simply assumed $M=2m_t$. We show the corresponding proton-level differential cross section $\dd\sigma(pp\to t\bar{t})/\dd m_{t\bar{t}}$ derived from the Green function as well as the B-W formula in Figure~\ref{fig:xs_pp_chi0t_GreenFunc}. By integrating over the ``resonance'' interval $m_{t\bar{t}}\in[M_{\chi_{t0}}-\Gamma,M_{\chi_{t0}}+\Gamma]=[343.43-2.84,343.43+2.84]$~GeV, one gets the cross section of $\sigma(pp\to\chi_{t0})_{\rm Green}=6.61~{\rm fb}$, which is compatible with $\sigma(pp\to\chi_{t0})_{\rm Projection}=10.0$~fb obtained through the projection method. 
Note that there is a displacement between the locations of the resonance peak predicted by the two methods, which could be due to the incomplete treatment of the $P$-wave Green function method adopted thus far. As demonstrated in Ref.~\cite{Iengo:2009ni}, the peak location of the $P$-wave spectrum predicted by the Green function method depends heavily on the relative velocity (and thus momentum) between the constituents $t$ and $\bar{t}$, while that of the $S$-wave spectrum is less sensitive, as can be seen from the results in Ref.~\cite{Lattanzi:2008qa}. Therefore, it is perhaps more favorable to formulate the $P$-wave Green function method in the momentum space, as has been proposed in Refs.~\cite{Fadin:1987wz,Fadin:1994pj} and carried out for the $S$-wave states in Ref.~\cite{Fuks:2024yjj}. On top of this, Ref.~\cite{deLima:2022joz} further points out that higher-order corrections to the $P$-wave Green function will induce a divergent term absent from the $S$-wave Green function, which, so far, cannot be dealt with using the known methods in the literature. As a result, it still remains to be explored the exact differential cross section predicted by the $P$-wave Green function method.

\setlength{\bibsep}{3pt}

\providecommand{\href}[2]{#2}\begingroup\raggedright\endgroup

\end{document}